\documentclass[%
 reprint, superscriptaddress,
 amsmath,amssymb,
 aps,
]{revtex4-2}
\pdfoutput=1 

\usepackage{graphicx}
\usepackage[dvipsnames]{xcolor}
\usepackage{subcaption}
\usepackage{dcolumn}
\usepackage{bm}
\usepackage{bbold}
\usepackage{makecell}
\usepackage{tikz-cd}
\usepackage{soul}
\usepackage{array}
\usepackage{booktabs}
\usepackage{physics}
\usepackage{amsmath,amsthm,amssymb,amsfonts,physics,siunitx,bbm,graphicx}
\usepackage{xfrac, hyperref}
\hypersetup{
    colorlinks,
    citecolor=blue,
    filecolor=blue,
    linkcolor=blue,
    urlcolor=blue
}
\usepackage{natbib}
\usepackage{mathtools}
\usepackage{comment}
\usepackage{stmaryrd}
\usepackage{braket}

\renewcommand{\arraystretch}{1.4}
\usepackage{svg}
\usepackage{subcaption}

\newtheorem*{result}{Result}
\newcommand{\ba}{\begin{eqnarray}}
\newcommand{\ea}{\end{eqnarray}}
\newcommand{\ban}{\begin{eqnarray*}}
\newcommand{\ean}{\end{eqnarray*}}
\newcommand{\one}{\mathbb{1}}

\newcommand{\chn}{\mathcal{E}}
\newcommand{\rf}{\gamma}
\newcommand{\rvp}{{\tilde{\chn}_{\rf}}}
\newcommand{\petz}{{\hat{\chn}_{\rf}}}
\newcommand{\pDW}{S^{\petz}_{\textbf{NQ}}}
\newcommand{\pDWW}{S^{\petz}_{\textbf{DW}}} 
\newcommand{\pSP}{S^{\petz}_{\textbf{SP}}}
\newcommand{\pQM}{S^{\petz}_{\textbf{QM}}}
\newcommand{\pCL}{S^{\rvp}_{\textbf{CL}}}
\newcommand{\TrB}{\text{Tr}_{B}}
\renewcommand{\tr}[1]{\text{\normalfont{Tr}}\!\left[#1\right]}
\newcommand{\swap}{U_{\leftrightarrowtriangle}}
\newcommand{\pswap}{U_{\!\curlywedgeuparrow}}
\newcommand{\sret}{S_{\mathcal{T}}^{\bar{\chn}_\rf}}

\newcommand{\ed}[1]{{#1}}
\newcommand{\edd}[1]{{#1}}

\allowdisplaybreaks

\begin{document}

\title{Quantum Bayesian Inference in Quasiprobability Representations}%

\author{Clive Cenxin Aw}
\affiliation{Centre for Quantum Technologies, National University of Singapore, 3 Science Drive 2, Singapore 117543}

\author{Kelvin Onggadinata}
\affiliation{Centre for Quantum Technologies, National University of Singapore, 3 Science Drive 2, Singapore 117543}

\author{Dagomir Kaszlikowski}
\affiliation{Centre for Quantum Technologies, National University of Singapore, 3 Science Drive 2, Singapore 117543}
\affiliation{Department of Physics, National University of Singapore, 2 Science Drive 3, Singapore 117542}

\author{Valerio Scarani}
\affiliation{Centre for Quantum Technologies, National University of Singapore, 3 Science Drive 2, Singapore 117543}
\affiliation{Department of Physics, National University of Singapore, 2 Science Drive 3, Singapore 117542}

\date{\today}

\begin{abstract}
Bayes' rule plays a crucial piece of logical inference in information and physical sciences alike. Its extension into the quantum regime has been the object of several recent works. These quantum versions of Bayes' rule have been expressed in the language of Hilbert spaces. In this paper, we derive the expression of the Petz recovery map within any quasiprobability representation, with explicit formulas for the two canonical choices of ``normal quasiprobability representations'' (which include Discrete Wigner representations) and of representations based on symmetric, informationally complete positive operator-valued measures (SIC-POVMs). By using the same mathematical syntax of (quasi-)stochastic matrices acting on (quasi-)stochastic vectors, the core difference in logical inference between classical and quantum theory is found in the manipulation of the reference prior rather than in the representation of the channel.
\end{abstract}

\maketitle

\section{Introduction}\label{intro}
Inference is a logical necessity in every science. In information theory and physics, the fundamentality of inference is particularly overt in notions of process reversibility and state recovery. Here, the most empirically applied and canonical approach is Bayes' rule:
\begin{equation} \label{bayes}
    \rvp(a|a')=\chn(a'|a)\frac{\rf(a)}{\tilde\rf(a')}.
\end{equation}
This relation gives us a recipe for obtaining various probability-theoretic objects \cite{watanabe65, watanabe55, jeffrey, jaynes_2003}. Of particular note, we may use it to obtain the ``reverse'' transition $\rvp$ for any given (i) the forward process or \textit{transformation} $\chn$, and (ii) the reference \textit{prior} $\rf$ on the input of said process. The \textit{posterior}, $\tilde\rf(a') = \sum_a \chn(a'|a) \rf(a)$, emerges from these two objects.

While the typical form of Bayes' rule works naturally for classical information theory, an extension to quantum theory requires some work. As one possible reason for this, notice that in a classical process $a\to a'$, the joint probability distribution $P(a,a')$ is routinely defined; and from this, one can compute marginal and conditional probabilities. By contrast, for a quantum process $\alpha\to\alpha'=\mathcal{E}(\alpha)$ where $\mathcal{E}$ is a completely positive trace preserving (CPTP) map, there is no elementary way to construct a state that takes both the input and the output into account. Various proposals have been presented over the years, and we refer to a very recent consolidating framework for all the references \cite{PF22}. In this context of finding a quantum theoretic Bayes' rule, a special role is played by the \textit{Petz recovery map} \cite{petz1,petz, wilde-recov}: 

\begin{eqnarray}\label{petz}
    \petz[\bullet] = \sqrt{\rf} \, \chn^\dagger\left[\frac{1}{\sqrt{\chn[\rf]}} \bullet \frac{1}{\sqrt{\chn[\rf]}}\right] \sqrt{\rf},
\end{eqnarray}
This recovery channel is defined for any CPTP map $\chn$ and a reference density operator $\rf$. Notably, when reference priors, input states and the channel share the same eigenbases, the Petz map reduces to the classical Bayes rule \cite{wilde-recov,wilde2011CItoQIreview,li-winter,PF22}. This and other properties pertaining to what may be called the ``conservation of divergences'' (which is what led to its conception) has built up this recovery map's reputation as the ``quantum Bayes' rule'' \cite{Leifer-Spekkens}; a reputation recently vindicated in an axiomatic approach \cite{petzisking2022axioms}. The Petz map construction appears also naturally in the definition of fluctuation theorems in thermodynamics \cite{kwon-kim,BS21,AwBS}.

Now, that said, it seems that what exactly makes the Petz map similar (or different) to the classical Bayesian update has not been as formalized as it could be. From an information-theoretical perspective, there are correspondences between the action of these recipes. Yet, we know that there are key regime-differences in the woodwork. This lack of formal comparison across these regimes is at least partially because the Petz map has thus far only been understood in terms of CPTP maps and density operators, living in a Hilbert space. Meanwhile, the classical Bayes rule exists as a stochastic matrix mapping stochastic vectors, living in a real vector space. While, as we noted, it is known that some CPTP maps correspond to the classical case, the reverse of this has yet to be done: writing the Petz maps in terms of matrices and vectors, without Hilbert space formalism.

In this paper, we attempt to close this gap by investigating the Petz map in \textit{quasiprobability representation} (QPR) \cite{ferrie2009framed, ferrie2011quasi}. This formalism provides a complete description of quantum theories while sharing the familiar mathematical equipment found in classical probability theory. The distinction is that \textit{quasiprobabilities} (or ``negative probabilities'') are generally necessary in the quantum case \cite{ferrie2008frame}. This negativity has been attributed as a resource for advantage in quantum computation \cite{veitch2012negative, howard2014contextuality, pashayan2015estimating}. As such, we seek to put the Petz map in the same formal habitat as that of classical Bayesian inversion and in an expression that is comparable to it. From there we may discuss the similarities, differences and interpretations wherever appropriate. We believe this work makes a formal step in understanding the essential distinctions between classical and quantum inference.

This paper is sectioned as follows. In Section \ref{BIReview}, we review features of Bayesian inference for classical and quantum transformations. In Section \ref{QPRReview}, we review the formalisms of QPR in quantum theory. Readers familiar with the formal content here may skim through these sections. In Section \ref{taskproper}, we work towards the key expression of the Petz map in QPR, stating relevant theorems along the way. In Section \ref{discussion}, we discuss consequent theoretical observations, contrasting notable formal features of the expression to the classical Bayesian update. In Section \ref{transitiongraphsection}, we introduce quasiprobabilistic ``transition graphs'' that can help visualize the implications of our results. Finally in Section \ref{conclude}, we summarize our findings and state some open lines of inquiry. 

\section{Classical \& Quantum Bayesian Inference} \label{BIReview}

In the context of classical mechanics and probability theory, a physical transformation can be expressed by conditional probabilities $\chn(a'|a)$ mapping probability distributions of inputs $p(a)$ to distributions of outputs $\tilde{p}(a') = \sum_{a} \chn(a'|a) p(a)$ residing in some given state space $A$ \footnote{Of course, one can have it that $a'$ and $a$ are defined in different state spaces $A$ and $A'$, but we can always take $A'' = A \cup A'$ and characterize the channel in this larger alphabet.}. This can be captured compactly by a stochastic matrix $S^\chn = \{\mathcal{E}(a'|a)\}$, mapping $v^p=\{p(a)\}$ to $v^{\tilde{p}}=\{\tilde{p}(a')\}$.

As already discussed, if we want to acquire a stochastically valid and logically sound ``reverse'' of this transformation $\chn$, we must invoke not only the channel in question but also a \textit{reference prior} $\rf$ on the input. This is essentially a pre-existing best guess of the inputs for which the Bayesian inverse is constructed. This process of acquiring $\rvp$ from $\chn$ and $\rf$ can be referred to as performing ``retrodiction'' (inference about the past, in contrast to prediction, inferring about the future) for $\chn$ on the prior $\rf$. Meanwhile, $S^{\rvp}v^{\tilde p}$ gives the ``retrodicted input'' given an observation $\tilde p$. It may also be referred to as the ``Bayesian update on $\rf$ given $\tilde p$''. 

For every each \textit{individual} transition , we may consult \eqref{bayes} for the corresponding retrodiction $a' \to a$. For the mapping of \textit{distributions}, it is more instructive to write the retrodiction map as a stochastic matrix: 
\begin{eqnarray}\label{bayesstoch}
    \pCL &=& D_\rf (S^\chn)^\text{T} D^{-1}_{\chn[\rf]}.\label{eq:cmret}
\end{eqnarray}
Here $D_p$ is a diagonal matrix with entries corresponding to some distribution $p$. 

As introduced in Section \ref{intro}, the counterpart to Bayes rule in quantum theory, is the Petz map \eqref{petz}. It is well-defined and CPTP for any full-rank $\chn[\rf]$ \footnote{This constraint also exists in the classical Bayes update and is likewise of no practical concern as one can always ensure that that $\rf$ is full-rank by adding some arbitrarily small weights into its spectrum and adding some arbitrarily small mapping probability in $\chn$ as well. These contributions can then be sent to zero on the recovered state.}. It may also be expressed as 
\begin{eqnarray}\label{petztriform}
    \petz= \mathcal{M}_{\rf^{1 \!/ 2}} \circ\chn^\dagger \circ \mathcal{M}_{\chn[\rf]^{-1 \!/ 2}},
\end{eqnarray}
where $\mathcal{M}_{\alpha^r}[\bullet] = \alpha^r \bullet \alpha^r$ for any density operator $\alpha$ and $r \in \mathbb{R}$, and $\chn^\dagger$ is the adjoint of $\chn$. This is the unique map for which 
\begin{equation}\label{eq:adj}
    \Tr(\chn[\rho] \sigma) = \Tr(\chn^\dagger[\sigma] \rho)
\end{equation}
for all self-adjoint $\rho,\sigma$.

Before continuing, it is important to stress that \textit{Bayesian inference is generically not inversion}. Inference is possible for any map, while inversion is only possible for invertible maps (information-preserving) -- and even then, the two operations are generally not the same, since the inverse of a map is generically not a valid map. In fact, it can be proved that inference and inversion coincide if and only if $S^{\chn}$ is a permutation (for the classical case), or $\chn$ is a unitary channel (in the quantum case) \cite{AwBS, PB22}. In general therefore, $\pCL S^{\chn} v^\rho \neq v^\rho$ and $\petz\circ\chn[\rho] \neq \rho$; although the reference state is recovered: $\pCL S^{\chn} v^\rf = v^\rf$ and $\petz\circ\chn[\rf] = \rf$ for all $\rf$.

\section{Quasiprobability representations} \label{QPRReview}
\subsection{Generalities}

We now move on to provide a brief review of the essential elements of QPRs for quantum theory. To map quantum theoretic objects acting on a $d$-dimensional Hilbert space to a QPR, the core is the choice of a {\it frame} $\{F_j\}_{j\in \Lambda}$, that is a set of $d\times d$ Hermitian operators spanning the Hermitian space equipped with Hilbert-Schmidt scalar product. The set of indices $\Lambda$ may be continuous; its minimal cardinality is $d^2$, and we shall assume such minimal frames for the remainder of this paper. Given a frame, one can always find a {\it dual frame} $\{G_j\}_{j\in \Lambda}$ such that
\begin{equation}\label{sumtraceproperty}
    \forall \, \alpha,\beta : \; \sum_{j}\tr{F_j \alpha}\tr{G_j \beta} = \tr{\alpha \beta}.
\end{equation}
In general, the dual is not unique given a frame. However, for a minimal basis, the frame and dual always satisfy the orthogonality relation $\tr{F_jG_k} = \delta_{jk}$.

Once a frame and its dual are set, all Hibert space objects are in one-to-one correspondence with an object in the QPR. \ed{A dictionary of recipes between the two frameworks are summarized in Table \ref{tab:QHM}. These recipes can be understood as category-theoretic relationships called ``functors'', mapping objects across the two formalisms which each live in their own separate category \cite{mac1998categories,carnap2002logical}}. From the normalisation of the state quasiprobability, $\sum_a v^\rho_a = 1$, it follows that the frame operators must satisfy $\sum_a F_a = \one$. Similarly, from the fact that each POVM $\{E_m\}$ must satisfy $\sum_m E_m = \one$, it follows that $\tr{G_j}=1$ for all $j\in\Lambda$ \footnote{It is the case, for SIC-POVM and Discrete Wigner representation, that $\tr{F_j} = 1/d$ for all $j \in \Lambda$. But this is not generally the case for all valid QPRs.}. As such, the QPR of any CPTP map $\mathcal{E}$ is a quasi-stochastic matrix $S^{\mathcal{E}}$ as defined in Table \ref{tab:QHM}, with entries $S^\mathcal{E}_{a'a} \in \mathbb{R}$, $a',a \in \Lambda$, s.t. $\forall a \; \sum_{a'} S^\mathcal{E}_{a'a} = 1$. \ed{The evolution of a state through a channel is then described by simple matrix multiplication: $\rho' = \mathcal{E}[\rho] \to v^{\rho'}_{a'} = \sum_{a}S^{\mathcal{E}}_{a'a}v^{\rho}_a$ \cite{ruzzi2000quantum}}. With a slight abuse of notation, for ease of correspondence with the classical formalism, we shall also denote the elements of the quasi-stochastic matrix as $S^\chn_{a'a} \equiv \chn(a'|a)$.

\onecolumngrid
\begin{center}
\begin{table}[t!]
\resizebox{0.7\textwidth}{!}{%
\begin{tabular}{|l|l|l|}
\hline
\textbf{Object} & \textbf{Hilbert space formalism} \hspace{0.5em}                     & \textbf{Quasiprobability formalism}    \hspace{0.5em}                                  \\ \hline
State          & $\rho=\sum_i \lambda_i\ket{\lambda_i}\bra{\lambda_i}$ & $v^{\rho} \; : \; v_a^{\rho}= \text{Tr}[\rho \,F_a]$          \\ \hline
POVM         & $\{E_m\,|\,E_m\geq 0 \,,\sum_m E_m = \one\}$                              & $\bar{v}^{m} \; : \; \bar{v}^m_{a'} = \text{Tr}[E_m \,G_{a'}]$                    \\ \hline
Unitary         & $\mathcal{U}[\bullet] = U\bullet U^{\dagger}, \; U U^\dagger = \one$ & $S^{\mathcal{U}}: S^{\mathcal{U}}_{a'a}= \tr{F_{a'}UG_{a}U^\dagger}$ \\ \hline
CP Maps  & $\mathcal{E}[\bullet] = \sum_l \kappa_l \bullet \kappa_l^\dagger \;  $  \; & $S^\mathcal{E}: \; S^\mathcal{E}_{a'a}=\text{Tr}[F_{a'} \mathcal{E}[G_a] ]$ \\ \hline
Born Rule       & $\text{Tr}[\rho E_m]$                            & $ v^{\rho} \cdot \bar{v}^m \in [0,1]$                           \\ \hline
Dimensionality \hspace{0.2em} & $\texttt{dim}[\mathbb{C}^{d}] =d $                    & $\texttt{dim}[\mathbb{R}^{d} \otimes \mathbb{R}^{d}] =d^2$ \\ \hline
\end{tabular}%
}
\centering
\caption{Dictionary of relationships between the Hilbert space and quasiprobability formalisms. $v^p_a = p(a)$ indicates the $a$-th entry in a $p$-distribution. $S^\chn_{a'a} = \chn(a'|a)$ indicates the entry on the $a'$-column and $a$-row of a matrix $S^\chn$. \ed{Frame and dual operators $F_i, G_j$ are defined within choices of representations as discussed in Sections \ref{ssect:NQPR} and \ref{ssect:SP}, with the most direct expressions found in \eqref{DWDFrame}, \eqref{DWFrame}, \eqref{SPFrame} and \eqref{SPconvert}.}}
    \label{tab:QHM}
\end{table}
\end{center}
\twocolumngrid
Some of the subsequent derivations apply generally to all representations; others are specific to one of the two canonical choices of QPR that we describe next.

\subsection{Normal quasiprobability representation}\label{ssect:NQPR}

The first class of representations are those for which the frame and dual frame operators are proportional to each other up to some scaling factor $c$, i.e., 
\begin{equation}\label{DWDFrame}
    G_j = cF_j 
\end{equation}
for all $j$. For minimal bases, the constant $c$ is equal to the Hilbert space dimension $d$. The class of representations satisfying this is known as {\it normal quasiprobability representation} (NQPR) \cite{zhu2016quasiprobability}.

An example of NQPR, and perhaps the most widely used representation, is the {\it discrete Wigner} (DW) representation \cite{wootters1987wigner,klimov2005discrete,gibbons2004discrete,gross2006hudson}, which is well-defined for prime dimension $d$ and composites of them. \ed{For odd-primes, the frame operators are defined as 
\begin{equation}\label{DWFrame}
    F_k = F_{r,s} =\frac{1}{d^2} \sum^{d-1}_{x,z=0} \omega^{sx-rz +\frac{xz}{2}}X^x Z^z\, ,
\end{equation}
where $k=(r,s)\in \mathbb{Z}_d\times \mathbb{Z}_d$, $\omega = e^{2\pi i/d} $ is the $d$-th root of unity, and $Z,X$ are generalized Pauli operators defined as $Z\ket{j} = \omega^j\ket{j}$, $X\ket{j} = \ket{j+1\mod d}$ with $\{\ket{j}\}_{j=0}^{d-1}$ as standard orthonormal basis. For a qubit system $(d=2)$,} the frame has a simple expression given by
\begin{equation}\label{canonDW}
    F_k =F_{r,s}=\frac{1}{4}\Big[\one +(-1)^{r} \sigma_x+(-1)^{s} \sigma_z+(-1)^{r+s} \sigma_y \Big],
\end{equation}
where $k=(r,s)\in \mathbb{Z}_2\times \mathbb{Z}_2$ \ed{and $\sigma_x,\sigma_y,\sigma_z$ are the familiar Pauli operators}. For composite $d = d_1\times d_2 \times \dots \times d_L$, where $d_1,d_2,\dots,d_L$ are primes, a tensor structure applies for the frame. That is, the frame operators decompose as
\begin{equation*}
        F_k = F_{k_1} \otimes F_{k_2} \otimes \dots \otimes F_{k_L},
\end{equation*}
where $k \to (k_1,k_2, \dots, k_L)$ with each $k_l = (r_l,s_l)\in\mathbb{Z}_{d_l}\times \mathbb{Z}_{d_l}$. This tensor structure is enjoyed by any NQPR and thus affords them an aesthetic benefit when dealing with composite states and purifications. 

\subsection{SIC-POVM representation}\label{ssect:SP}

Under NQPR, negativity can be found in states, POVM elements, and transformations alike. Symmetric, informationally complete, positive operator-valued measure (SIC-POVM) representations seek to avoid this by ensuring that all state vectors are positive \cite{appleby2017introducing,kiktenko2020probability}. Negativity features are thus consolidated into the transformations and POVMs 

For $d$-dimensional Hilbert space, a SIC-POVM is defined as a set of sub-normalized rank-1 projectors $\{\frac{1}{d}\Pi_j\}_{j=1}^{d^2}$, $\Pi_j =\ketbra{\psi_j}$, such that the elements have equal pairwise Hilbert-Schmidt inner product:
\begin{equation}\label{SPFrame}
    \tr{\Pi_j^\dagger\Pi_k} = |\!\braket{\psi_{j}|\psi_{k}}\!|^2 = \frac{d\delta_{jk}+1}{d+1}.
\end{equation}
The solution to the vectors of SIC-POVM have been found for vast number of dimensions (see \cite{debrota_qbism} for the list), and is believed to exist for all \cite{appleby2013galois}. Since the set is informationally complete (i.e. it forms a basis) we can use it as the definition of the SIC-POVM representation's frame $\{F_j = \frac{1}{d}\Pi_j\}$. From the orthogonality relation, it can be easily deduced that the dual frame is given by
\begin{equation} \label{SPconvert}
    G_j = d(d+1)F_j - \one = (d+1)\Pi_{j} - \one.
\end{equation}
\ed{For one qubit, the canonical choice is the tetrahedron
\begin{equation}\label{canonSP}
    F_j = \frac{1}{4}\left[\one + \vec{v}_j\cdot \vec{\sigma}\right],
\end{equation} where $\vec{\sigma} = (\sigma_x,\sigma_y,\sigma_z)$ and $\vec{v}_0=\frac{1}{\sqrt{3}}(1,-1,1)$, $\vec{v}_1=\frac{1}{\sqrt{3}}(-1,1,1)$, $\vec{v}_2=\frac{1}{\sqrt{3}}(1,1,-1)$, $\vec{v}_3=\frac{1}{\sqrt{3}}(-1,-1,-1)$.}

\section{The Petz Map in Quasiprobability Formalisms}\label{taskproper}
Now, our task is to express the Petz recovery map in its QPR, which we denote as $S^\mathcal{\petz}$. This obviously can be done by invoking the relationship between maps in Table \ref{tab:QHM} and then connecting it with \eqref{petz}. This gives:
\begin{eqnarray}\label{petztrivialapproach}
     S^{\petz}_{aa'} &=& \tr{F_{a}\sqrt{\rf} \chn^\dagger\left[\frac{1}{\sqrt{\chn[\rf]}} G_{a'} \frac{1}{\sqrt{\chn[\rf]}}\right] \!\sqrt{\rf}\,}.
\end{eqnarray}
But, of course, this affords us no new insight. We are still relying entirely on the Hilbert space formalism. Nothing novel can be said in comparison to classical Bayesian inference as found in \eqref{bayesstoch}. Our specific task is as illustrated in FIG. \ref{commdiag}: write the Petz in a way that \textit{only quasiprobability-theoretic objects} (quasi-stochastic vectors, matrices and frames) \textit{are required}. 

\begin{figure}[h]
    \centering
\begin{tikzcd}[sep=4.5em]
\chn, \rf \arrow[r, "\textbf{QPR}" description] \arrow[d, "\textbf{PETZ}" description] & S^{\chn}, v^\rf \arrow[d, "\fbox{\textbf{?}}" description] \\
\hat{\chn}_\rf \arrow[r, "\textbf{QPR}" description]                              & S^{\hat{\chn}_\rf}                          
\end{tikzcd}
    \caption{\ed{A commutativity diagram illustrating the main task of this work: the protocol ``\textbf{?}'', that is to be executable solely within the QPR framework.}}
    \label{commdiag}
\end{figure}

The naive guess that $S^{\petz}$ could be obtained by grafting the quasiprobabilistic formalism onto the classical Bayesian inverse \eqref{bayesstoch} is easily dismissed: the $\pCL$ obtained by such a recipe is in general not a valid map in QPR (see Appendix \ref{Q-is-not-C} for explicit counterexamples) as it results in measurement outcome statistics that are out of bounds. That is, 
\begin{equation*}
    \exists(\chn, \rf, \rho, m ): (\pCL v^\rho) \cdot \bar{v}^m \notin [0,1].
\end{equation*}
\ed{Rather, let's start by noticing that the recipe for channels in Table \ref{tab:QHM} and condition \eqref{sumtraceproperty} imply the concatenation
\begin{equation}\label{concat} S^{\chn} S^{\mathcal{F}} = S^{\chn \circ \mathcal{F}}\end{equation} for two channels $\chn, \mathcal{F}$. Thus the Petz map  \eqref{petztriform} is represented by}
\begin{equation}\label{prelim}
    S^{\petz} = M_{\rf^{1 \!/ 2}} \, \big(S^{\chn^\dagger}\big) \, M_{\chn[\rf]^{-1 \!/ 2}}
\end{equation}
\ed{with
\begin{eqnarray}\label{defM}
    M_{\alpha^{r}}&\coloneqq& S^{\alpha^r\,\bullet\, \alpha^r}
\end{eqnarray}
that is}
\begin{eqnarray}\label{m-iso}
     (M_{\alpha^{r}})_{a'a} = \tr{F_{a'} \alpha^r G_{a} \alpha^r}.
\end{eqnarray}
Now, it is crucial for our goals that all objects entering \eqref{prelim} can be constructed within the quasiprobability formalism.

As a first check, we notice that all the entries of the matrices $M_{\alpha^{r}}$ are real. Indeed, one can rewrite $(M_{\alpha^{r}})_{a'a} = \tr{\mathcal{F}^r_{a'}\mathcal{G}^r_a}$ with $\mathcal{F}^r_a=\alpha^{r/2} F_{a} \alpha^{r/2}$ and $\mathcal{G}^r_a=\alpha^{r/2} G_{a} \alpha^{r/2}$. These are Hermitian operators, and so $\tr{\mathcal{F}^r_{a'}\mathcal{G}^r_a}=\frac{1}{2}\tr{\{\mathcal{F}^r_{a'},\mathcal{G}^r_a\}}$ is real.

Next, we show that $M_{\alpha}$ can be expressed using the quasiprobability representation of the state $\alpha = \sum_x v^\alpha_x G_x$. Indeed,
\begin{eqnarray}
    (M_{\alpha})_{a'a} &=& \tr{F_{a'} \alpha G_{a} \alpha} \nonumber\\
    &=& \sum_{xy} v^\alpha_x v^\alpha_y \tr{F_{a'} G_x G_{a} G_y}\label{interm}\\
    &\coloneqq& \label{M-entry} \sum_{xy} v^\alpha_x v^\alpha_y \xi_{a'xay}, \label{closedform}
\end{eqnarray}
where the
\begin{equation}\label{eq:structure}
    \xi_{pqrs} = \tr{F_p G_q G_r G_s}
\end{equation}
are referred to as {\it structure coefficients}. \ed{These same coefficients have appeared in a recent work on the DW representation of maps \cite{braasch2020transition} (see Appendix \ref{WoottDeriv}). The $M_\alpha$ are real, positive-semi-definite matrices with a unit trace: $M_\alpha \geq 0$ and $\tr{M_{\alpha}}= 1$. For NQPRs, they are Hermitian, while for SIC-POVMs they are generically not symmetric and thus not Hermitian (see Appendix \ref{propertiesofMalpha}.

Having expressed $M_\alpha$ in the QPR formalism, one can finally prove that
\begin{equation}
    M_{\alpha^{r}} = M_{\alpha}^r
\end{equation} holds for any $r \in \mathbb{Q}$ (that is, for any rational number $r$, see Appendix \ref{powertopower} for the proof).} In particular, the $M_{\alpha^r}$ for $r=\pm\frac{1}{2}$ that are needed in \eqref{prelim} can be constructed from the quasiprobability representation of states by first computing $M_\alpha$ \eqref{closedform}, then taking the suitable roots.


%

In summary, we have obtained our main result:

\begin{result}
The Petz map in any QPR reads
\begin{equation}\label{main}
    \pQM = M_{\rf}^{1 \!/ 2} \, \big(S^{\chn^\dagger}\big) \, M_{\chn[\rf]}^{-1 \!/ 2},
\end{equation}
where 
\begin{eqnarray*}
    (M_{\rf})_{a'a} &=& \sum_{xy} v^\rf_x v^\rf_y \, \xi_{a'xay}, \\
    \big( M_{\chn[\rf]} \big)_{a'a} &=& \sum_{xy} (S^{\chn} v^\rf)_x (S^{\chn} v^\rf)_y \, \xi_{a'xay}.
\end{eqnarray*} and $\xi_{pqrs} = \tr{F_p G_q G_r G_s}$ are structure coefficients determined by the specific QPR. Everything is expressed exclusively in the quasiprobabilistic formalism: no knowledge of Hilbert space renditions of the quantum channel or reference state is required. 
\end{result}

For the two canonical choices of QPR introduced above, we prove in Appendix \ref{adjoints} that
\begin{eqnarray}
    \text{\textbf{NQPR}}\,: \; S^{\chn^\dagger}_{\textbf{NQ}} \! &=& \! (S^{\chn})^\text{T}, \\ 
    \text{\textbf{SIC-POVM}}\,: \; S^{\chn^\dagger}_{\textbf{SP}} \! &=& \! (S^{\chn})^\text{T} + J_{\chn}, \label{SPadjoint}
\end{eqnarray}
where $(J_{\chn})_{ij}= \frac{1}{d}(\sum_a \chn(j|a)-1)$; whence explicitly
\begin{eqnarray}
    \pDW &=& M_{\rf}^{1 \!/ 2} (S^{\chn})^\text{T} M_{\chn[\rf]}^{-1 \!/ 2}, \label{petzDW} \\ 
    \pSP &=& M_{\rf}^{1 \!/ 2} \left[(S^{\chn})^\text{T}+J_{\chn}\right] M_{\chn[\rf]}^{-1 \!/ 2}. \label{petzSIC}
\end{eqnarray}
Since the QPR of unital maps (i.e. $\chn[\one] = \one$) are quasi-bistochastic matrices (that is, $\sum_a \chn(j|a) = 1$ for all $j$), for such maps $J_{\chn}$ vanishes and the expressions for NQPR and SIC-POVM representations are formally identical.
\vspace{1em}
\renewcommand\cellgape{\Gape[4pt]}
\renewcommand{\arraystretch}{1.6}

\ed{\section{Comparing classical and quantum retrodiction}\label{discussion}
\subsection{Main comparison}\label{discussion-A}
We are finally in a position to compare classical and quantum Bayesian inference. Having found \eqref{main}, it is easy to notice that the classical Bayes' rule \eqref{bayesstoch} can be rewritten in the same form: 
\begin{equation} \label{bayesstoch-amended}
    \pCL =  (D_{\rf}^2)^{1 \!/ 2} \, \big(S^{\chn^\dagger}\big) \, (D_{\chn[\rf]}^2)^{-1 \!/ 2}
\end{equation}
because $(S^\chn)^\text{T} = S^{\chn^\dag}$ for classical channels (see Appendix \ref{classicaladjoint}) and $(D_{\rf})_{ij}=v^{\rf}_{i}\delta_{ij}$. In other words, classical Bayesian inference hides the fact \textit{that the central matrix is an adjoint, and that the left and right matrices should be seen as square roots of more fundamental matrices $X_\rf$ and $X_{\chn[\rf]}$}. This is the \textit{common form} of classical and quantum Bayesian inference that emerges from using QPRs.

Let us now study the \textit{differences} between the two theories. There is of course the starting point: for a given system dimension $d$,  in classical theory the $v^\gamma$ are $d$-component probability vectors, while in a minimal QPR of quantum theory they are $d^2$-component quasiprobability vectors. If we leave this aside and we focus only on Bayesian inference, the \textit{formal difference} appears in the matrices $X_\rf$ ($D_\rf^2$ for classical, $M_\rf$ for quantum). Both can be written
\begin{equation}\label{bayes-amend}
    (X_{\rf})_{ij} = \sum_{x,y=1}^{d^2} v^\rf_x v^\rf_y \xi_{ixjy}\,;
\end{equation} but while the $M_\rf$ of quantum theory has $\xi_{ixjy}=\tr{F_i G_x G_j G_y}$, the $D^2_\rf$ of classical theory is such that $(D^2_{\rf})_{ij}=({v^\rf_i})^2\delta_{ij}$ i.e.
\begin{equation}\label{claxprej}
\xi_{ixjy} = \delta_{ix} \delta_{jy} \delta_{ij} \quad\quad \textrm{[Classical]}\,.   
\end{equation}
This comparison is summarized in Table \ref{tab:compares}.}

\begin{center}
\begin{table}[h!]
\resizebox{0.41\textwidth}{!}{%
\begin{tabular}{|c|c|c|}
\multicolumn{3}{c}{\textbf{Bayesian Inference in Theory $\mathcal{T}$}} \\ 
\multicolumn{3}{c}{\makecell{$ \sret = X_{\rf}^{1 \!/ 2} \big(S^{\chn^\dagger}\big) X_{\chn[\rf]}^{-1 \!/ 2} $  }} \\ 
\multicolumn{3}{c}{\makecell{$ (X_{\rf})_{j} = \sum_{xy} v^\rf_x v^\rf_y \xi_{ixjy} $                  \vspace{0.5em}}} \\ 
\hline
\hspace{4pt} \textbf{Object} \hspace{4pt}    &  \hspace{14pt} \textbf{$\mathcal{T}$: Quantum}  \hspace{14pt}     &\hspace{6pt} \textbf{$\mathcal{T}$: Classical}  \hspace{6pt}  \\ \hline
$S^{\chn^\dagger}$    & \makecell{$\textbf{NQ}: (S^{\chn})^\text{T}$ \\  $\textbf{SP}: (S^{\chn})^\text{T} + J_{\chn}$}
                    & $(S^{\chn})^\text{T}$     \\ \hline
$\xi_{ixjy}$        & $\tr{F_i G_x G_j G_y}$                 & $\delta_{ix} \delta_{jy} \delta_{ij}$ \\ \hline
\end{tabular}%
}
\centering
\caption{Retrodiction maps for classical probabilities [$\sret \to \pCL$, Eq.~\eqref{bayesstoch} or \eqref{bayesstoch-amended}] and quantum quasiprobabilities [$\sret \to S^{\petz}_\textbf{QM}$, Eq.~\eqref{main}]. \edd{The law of transformation of the central object is the same both in classical and in quantum theory; while one needs the structure coefficients to obtain the $X_\gamma$ from the $v_\gamma$.}}
    \label{tab:compares}
\end{table}
\end{center}

\ed{\subsection{Related remarks}

Previously, we mentioned that $\pQM$ is not naively equal to $\pCL$. We can now be more precise about this difference. First, we notice that no frame satisfies $\tr{F_i G_x G_j G_y} = \delta_{ix} \delta_{jy} \delta_{ij}$. If this were the case, $M_\gamma$ would be diagonal for all $\gamma$; in fact, all conceivable quantum channels would become trivial. Here is the proof: by extending on \cite{braasch2020transition} (see Appendix \ref{WoottDeriv}), in any QPR every entry of $S^\chn$ can be expressed as
\begin{equation} \label{eq:altse}
        S^\chn_{ij} = \sum_{xyl} \tr{F_x \kappa_l} \tr{F_y \kappa_l^\dag} \tr{F_i G_x G_j G_y}
\end{equation}
where $\{\kappa_l\}_l$ \edd{defines a Kraus representation of $\chn$ \footnote{\edd{Our main claim, summarized in Table \ref{tab:compares}, is that, in deriving the retrodiction \textit{inside a QPR}, the structure coefficients are used only to obtain the $X_\gamma$ from the $v_\gamma$. Eq.~\eqref{eq:altse} shows that one can use the structure coefficients to obtain the QPR $S^\chn$ \textit{from the objects of the Hilbert space}: thus, this fact does not contradict our main claim.}}.} Thus, a hypothetical frame satisfying \eqref{claxprej} would have
\begin{align*}
    S^\chn_{ij} &= \sum_{xyl} \tr{F_x \kappa_l} \tr{F_y \kappa_l^\dag} \delta_{ix} \delta_{jy} \delta_{ij} \\
    &= \delta_{ij} \sum_l \tr{F_i \kappa_l} \tr{F_j \kappa_l^\dag}=\delta_{ij}\,, 
\end{align*}
where the last equality comes from the fact that $S^\chn$ must be quasistochastic. In other words, we would have $S^\chn = \one_{d^2}$ for every $\chn$, which is absurd \footnote{For NQPR and SIC-POVMs, the proof is even simpler using the fact that $\sum_x G_x = d \one$: indeed, in this case we have $\sum_{xy} \tr{F_i G_x G_j G_y}=d^2 \tr{F_i G_j}$; but $\sum_{xy}\delta_{ix} \delta_{jy} \delta_{ij}=\delta_{ij}$, and therefore \eqref{claxprej} can hold only if $\quad d^2 \tr{F_i G_j} = \delta_{ij}$ i.e.~$d^2=1$.}. We conjecture that the only way to obtain $\tr{F_i G_x G_j G_y} = \delta_{ix} \delta_{jy} \delta_{ij}$ with $d\times d$ matrices is to set $F_i=G_i=\ket{i}\bra{i}$ for $1\leq i\leq d$ and $\braket{i|j}=\delta_{ij}$; but this defines a projective measurement, it has only $d$ elements and is certainly not a frame.

Next, one might conjecture that $(\pQM v^{\rho}) \cdot \bar{v}^m = (\pCL v^{\rho}) \cdot \bar{v}^m$ holds for classical processes: that is, processes in  which $\rho, \rf, \chn[\rf], E_m$ all have the same eigenbasis and no coherence appears. While our numerical exploration suggests that this is the case for some specific scenarios, it is certainly false in general, even for simple examples (see Appendix \ref{CounterExamplesCQ}). }


A last question is whether the quantum Bayes rule can be written as \begin{equation}
    \petz(a|a') = f\Big(\{\chn(a'|a)\}, \,\{\rf(a)\}\Big)
\end{equation} in analogy to Eq.~\eqref{bayes}. The answer is: \textit{in principle, yes}, because the square roots of $M_{\rf}$ and $M_{\chn[\rf]}$ are certainly functions of the $\chn(a'|a)$ and the $\rf(a)$. However, writing down this expression \textit{in practice} requires the explicit expressions. Even for the simplest quantum case (the qubit), in general one would have to find the roots of a quartic characteristic equation. 

\section{Visualizing Quantum Inference via QPR}\label{transitiongraphsection}
\subsection{Introducing Transition Graphs}
A notable advantage of stochastic maps is their ease of visualization. One can draw what might be called ``transition graphs'', where transition between $a_i$ to $a_{j}'$ are depicted by arrows going from the former to the latter. The probabiltiy weights on these transitions may be then depicted by a number or by a colour function. These kinds of graphs are not straightforward to write for the standard Hilbert space formalism. This is simply due to the use of complex terms, probability amplitudes and the plurality of possible basis choices. With QPR, we can illustrate transformations and their quantum Bayesian inverses with transition graphs just as we would for classical stochastic channels, albeit with the added task of depicting negativity in these transitions. 

In Appendix \ref{app:graphs} and this section, we consider some choices of $\chn$ that give rise to $S^\chn$ 
and their retrodictions $\pDWW$ and $\pSP$. These are then depicted as transition graphs. We have chosen to include, in particular, a Half-SWAP channel with a $\ketbra{1}{1}$ ancilla to visually illustrate and explore the properties of quantum retrodiction. Other transformations are also noted in passing with their graphs and expressions consolidated in Appendix \ref{app:graphs}. Before these, we note some illustrative elements of these figures. 

Firstly, with transition arrows we depict negative (positive) quasiprobabilities with cooler (warmer) shades. Furthermore, these negative (positive) arrows will be drawn with dashed (solid) lines. A colour legend is included in FIG. \ref{fig:colourbar}. 

Secondly, in order to get a \ed{sense} of how irreversible a \textit{forward} map is and which states it tends to erase toward, we add coloured ``bubbles'' around the \textit{output} side (denoted $\{a'_j\}$) of every graph for a given $S^\chn$. The intensity and colour of the bubbles are weighted according quasiprobability distribution of the state $\chn[\one/d]$. Hence, one should expect that these bubbles are coloured uniformly for all unital maps \ed{(and thus, for all reversible maps too)}. \ed{Visually speaking, the most irreversible maps would be those for which these coloured bubbles correspond exactly to the colour of the transition arrows that are drawn toward them (see FIG. \ref{fig:SWAP} for an example of this).}\footnote{The equivalence in colour is a statement of irreversibility since it implies $\chn(a'|a) = \hat{\rf}(a')$. Thus $\sum_{a}\chn(a'|a)q(a) = \hat{\rf}(a')$ for all $q(a)$, which is just to say the channel irreversibily erases all information about the input.} 

Thirdly, a similar feature is added for the \textit{retrodictive} transition graphs, drawn for $S^{\petz}$ matrices. Crucial for understanding the Bayesian inverse is the reference prior. Hence, for Bayesian inverting transition graphs we add coloured bubbles on the \textit{input} (denoted $\{a_j\}$, that is, the input of the \textit{forward} map) side of the graph, weighted according to the distribution of $\rf$. \ed{For simplicity, we describe channels acting on qubits and use the most canonical choices of frames: \eqref{canonDW} for DW (with $r,s$ starting from $0$) and \eqref{canonSP} for SIC-POVM representations. We employ these frames for all the numerics found in this paper.} 

\subsection{Fully Reversible \& Fully Irreversible}\label{fullyrevfullirrev}

As depicted in Figures \ref{fig:perm} and \ref{fig:hadaarbi} (found in Appendix \ref{app:graphs}), we observe the provable property that $S^{\hat{\mathcal{U}}_\rf} = S^{\hat{\mathcal{U}}} = (S^{\mathcal{U}})^\text{T}$, for unitary channels $\mathcal{U}$. The Bayesian inverses simply reflect the transition trajectories back, doing so with equal probability and negativity and regardless of what reference prior is chosen. More interesting features occur for non-unitary channels. We may write any CPTP map as a dilation defined by a global unitary $U$ acting on an extended state space $\mathcal{H}_A \otimes \mathcal{H}_B$ for which the input system $\bullet_A$ and an environment or ancilla $\beta_B$ is defined: 
\begin{equation}\label{dil}
    \chn[\bullet] = \TrB[U \bullet \otimes \beta \, U^\dagger].
\end{equation}
We stick to the case where both the target and the ancilla are qubits. Arbitrary qubits may be written as:
\begin{equation}\label{arbiqubit}
    \beta(\omega,\theta,\phi) = \sin^2 (\omega) \ketbra{\psi} +\cos^2 (\omega) \ketbra{\psi^\bot}.
\end{equation}
Where $\ket{\psi} = \cos(\theta/2) \ket{0} + e^{i\phi} \sin(\theta/2) \ket{1}$ and $\ket{\psi^\bot} = e^{-i\phi} \sin(\theta/2) \ket{0} + \cos(\theta/2)  \ket{1}$. In maximal contrast to unitary channels, one may consider a quantum total erasure channel. This is simply a kind of replacement map where a Full-SWAP \eqref{swapeq} acts on a qubit and an ancilla and we trace out the environment. The Bayesian inverse of such quantum channels follow their classical counterparts: they erase back to reference prior \cite{AwBS}. Since the channel is totally irreversible, the quantum Bayes rule simply reverts our inference to our best guess about the initial state (illustrated by FIG. \ref{fig:SWAP}). 

\subsection{Liminally (Ir)reversible}\label{PetzforHSWAP}

For a more conceptually involved and instructive scenario, we consider the Half-SWAP $\pswap$, which may be represented in the computational basis as: 
\begin{eqnarray}
\pswap \; \hat{=} \; \frac{1}{\sqrt{2}}  
\begin{pmatrix}
 \sqrt{2} & 0 & 0 & 0 \\
 0 & 1 & 1 & 0 \\
 0 & 1 & -1 & 0 \\
 0 & 0 & 0 & \sqrt{2} \\
\end{pmatrix} 
\end{eqnarray}

\begin{figure}[h!] 
\begin{subfigure}{0.45\textwidth}
\centering
         \includegraphics[width=0.85\linewidth]{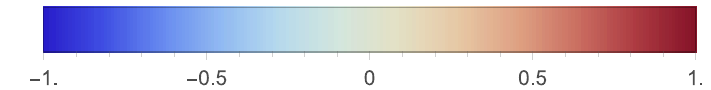}
         \caption{Colour Legend for \ed{Quasiprobabilities}} \label{fig:colourbar}
\end{subfigure} 
\begin{subfigure}{0.45\textwidth}
\centering
         \includegraphics[width=0.95\linewidth]{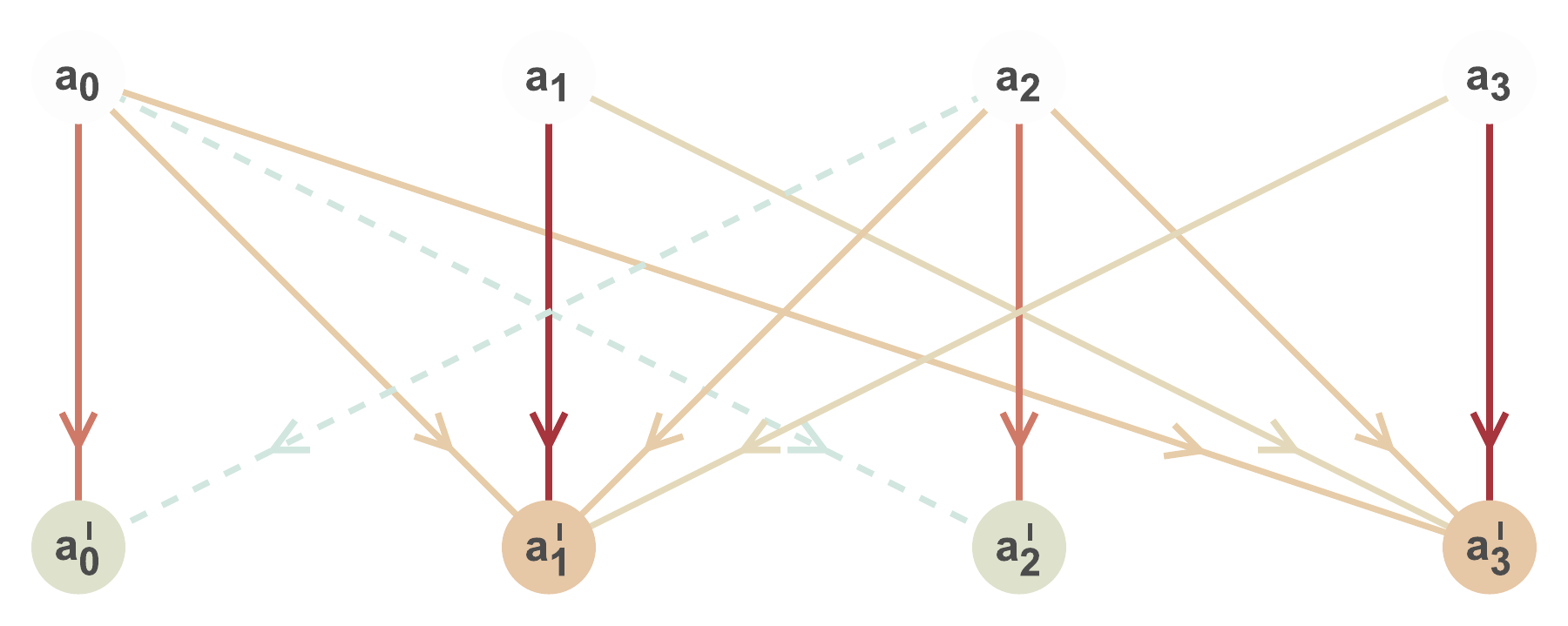}
         \caption{$S^{\chn}_\textbf{DW}$ for $\chn[\bullet]=\TrB[\pswap \bullet \otimes \ketbra{1} \, \pswap^\dagger]$} \label{fig:psf}
\end{subfigure} 
\begin{subfigure}{0.45\textwidth}
\centering
         \includegraphics[width=0.95\linewidth]{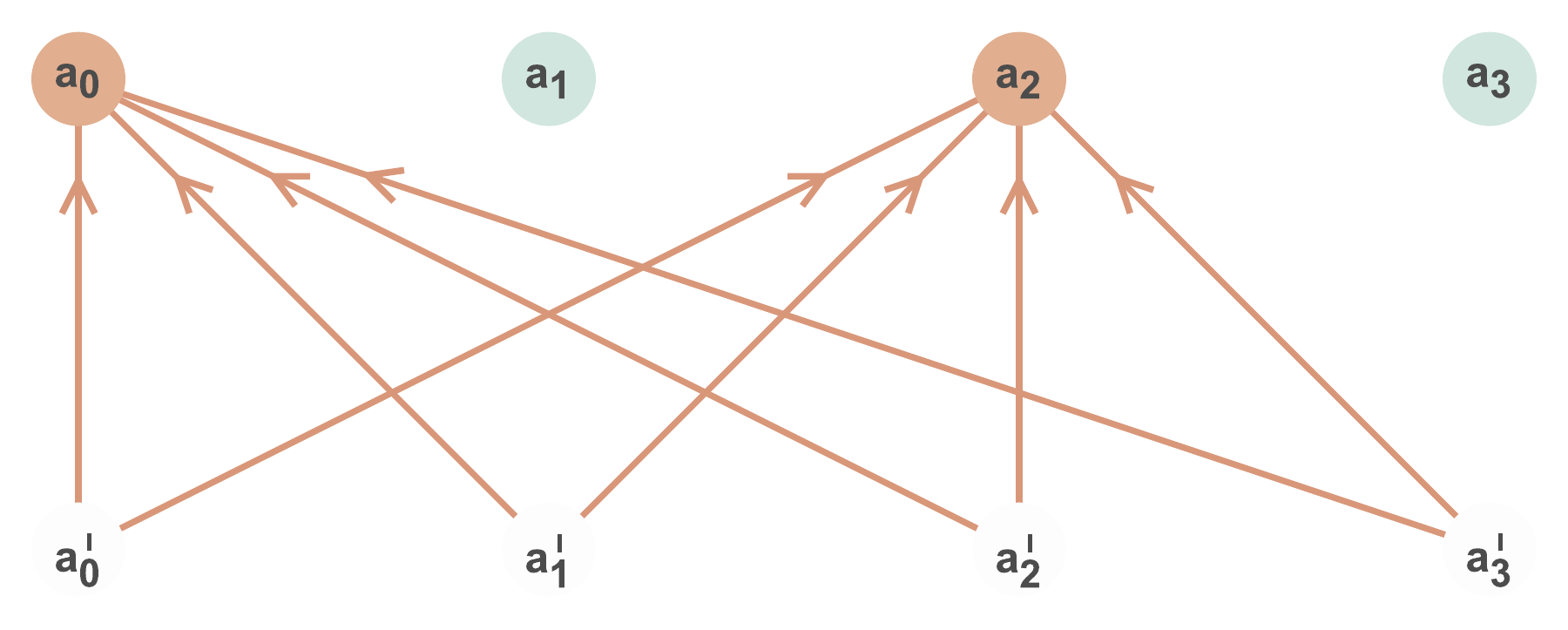}
         \caption{$\pDWW$ for $\pswap, \beta =  \ketbra{1}, \rf =  \ketbra{0}$} \label{fig:ps0}
\end{subfigure} 
\begin{subfigure}{0.45\textwidth}
\centering
         \includegraphics[width=0.95\linewidth]{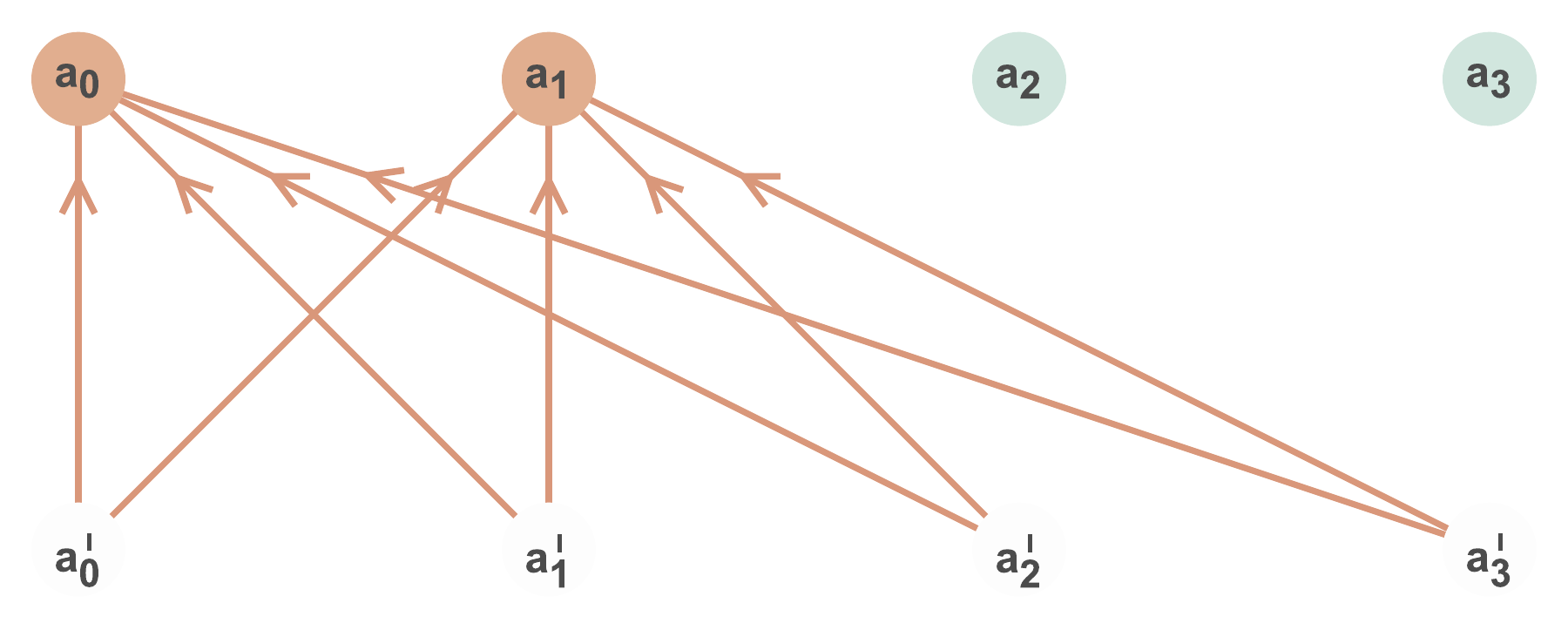}
         \caption{$\pDWW$ for $\pswap, \beta =  \ketbra{1}, \rf =  \ketbra{+}$} \label{fig:ps+}
\end{subfigure} 
\begin{subfigure}{0.45\textwidth}
\centering
         \includegraphics[width=0.95\linewidth]{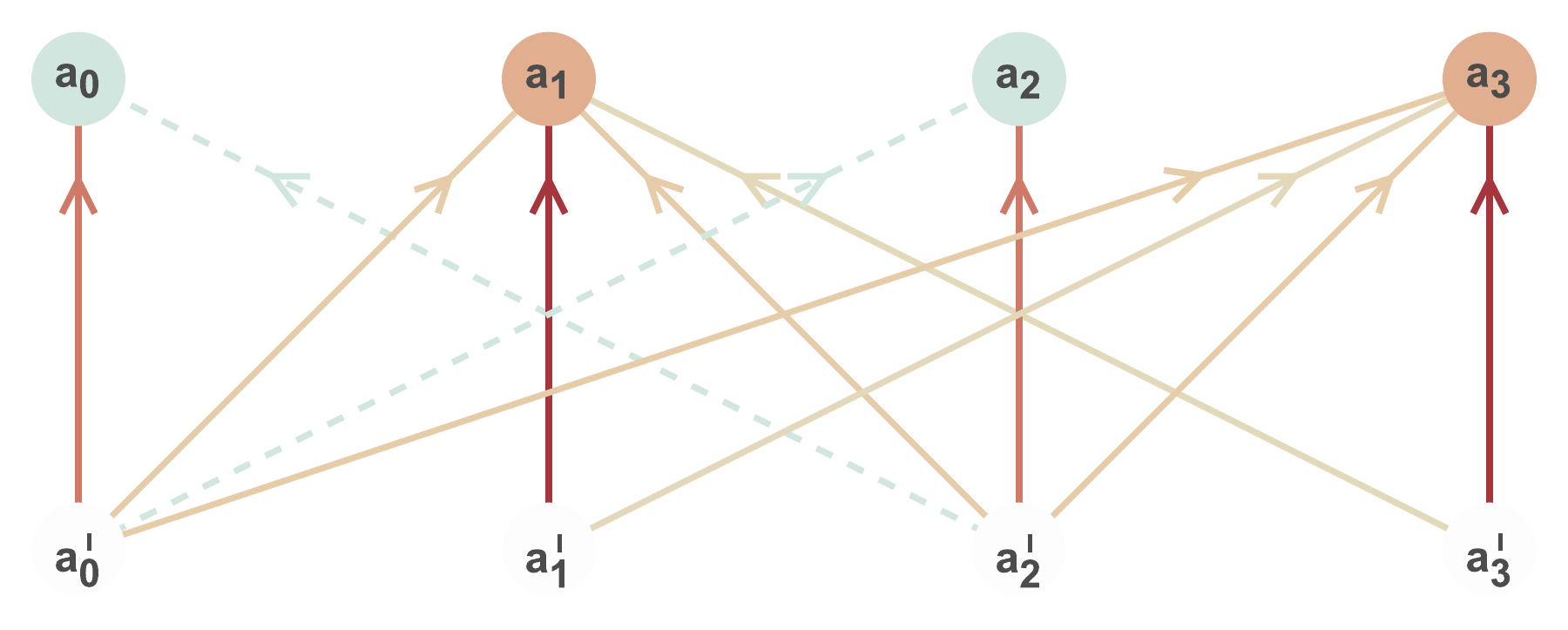}
         \caption{$\pDWW$ for $\pswap, \beta =  \ketbra{1}, \rf =  \ketbra{1}$} \label{fig:ps1}
\end{subfigure} 
\begin{subfigure}{0.45\textwidth}
\centering
         \includegraphics[width=0.95\linewidth]{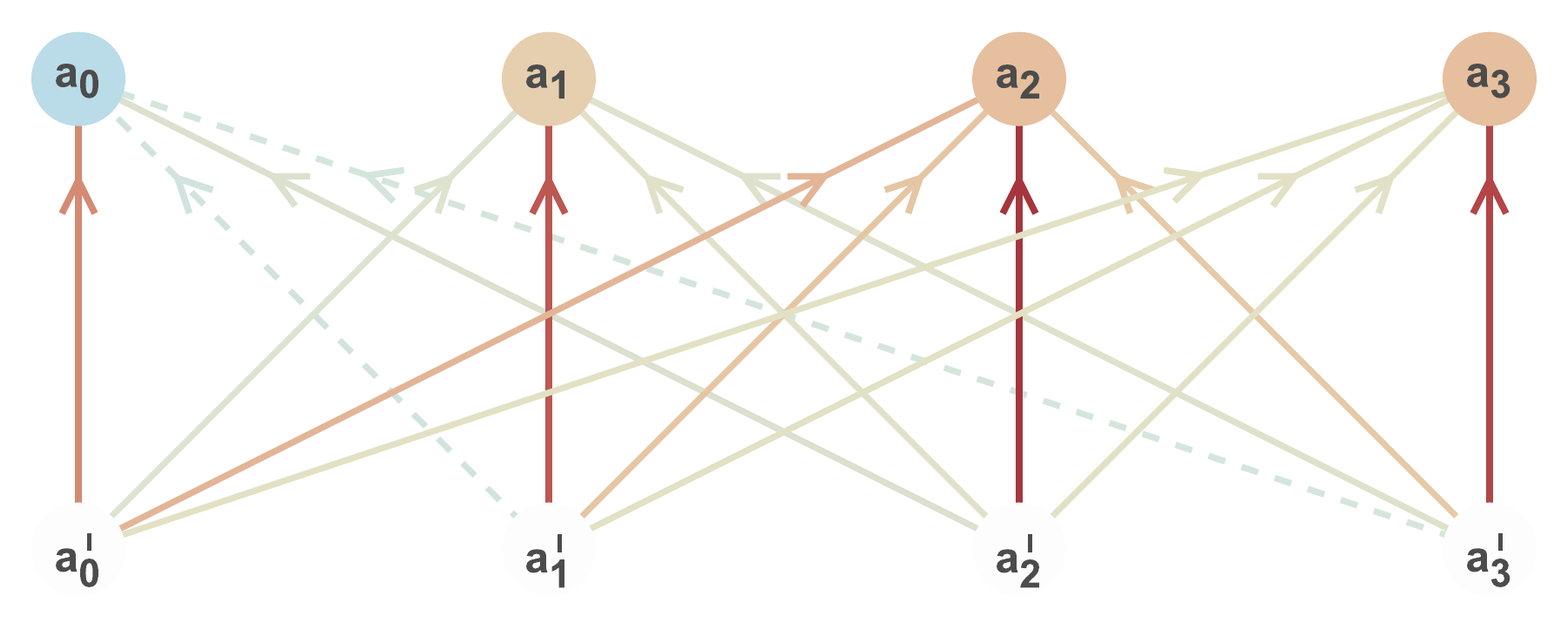}
         \caption{$\pDWW$ for $\pswap, \beta =  \ketbra{1}, \rf(\frac{\pi }{16},\frac{\pi }{5},\frac{\pi }{3})$} as per \eqref{arbiqubit} \label{fig:psarbi}
\end{subfigure} 
     \caption{Transition Graphs for a ``Half-SWAP'' with $\ketbra{1}$, and various retrodictions with a range of reference priors.}
     \label{fig:PSWAP}
\end{figure} 

As depicted in Figures \ref{fig:PSWAP}, we have the forward and retrodictive transition graphs for a channel given by $\chn[\bullet]=\TrB[\pswap \bullet \otimes \ketbra{1} \, \pswap^\dagger]$. To understand the retrodictive action given by the Petz, we can gain some intuitions \ed{by} writing out these mappings:
\begin{eqnarray*}
    \ket{01}  &\overset{\pswap}{\xrightarrow{\hspace*{0.45cm}}}& \frac{1}{\sqrt{2}}\big( \ket{01}+\ket{10} \big) \overset{\text{Tr}_B}{\xrightarrow{\hspace*{0.45cm}}} \fbox{$\displaystyle{\frac{1}{2}\one}$} \\ 
        \ket{+1} &\overset{\pswap}{\xrightarrow{\hspace*{0.45cm}}}& \frac{1}{2}\ket{11} + \frac{1}{\sqrt{2}}\big( \ket{01}+\ket{10} \big) \\ &\overset{\text{Tr}_B}{\xrightarrow{\hspace*{0.45cm}}}& \fbox{ $\displaystyle{{\frac{1}{4}} \ketbra{0}+\frac{3}{4}\ketbra{1}+ \frac{1}{2\sqrt{2}}\big(\ketbra{1}{0}+\ketbra{0}{1} \big)}$} \\
    \ket{11} &\overset{\pswap}{\xrightarrow{\hspace*{0.45cm}}}& \ket{11} \overset{\text{Tr}_B}{\xrightarrow{\hspace*{0.45cm}}} \fbox{$\displaystyle{\ketbra{1}}$} \\
\end{eqnarray*} 

We see that if the reference state is $\rf = \ketbra{0}$ or $\ketbra{+}$, then any state is compatible to its output (they are unambiguously full rank in $\mathbb{C}^2$). Hence, the Petz Map erases all (output) states back to the reference, in full consistency with the earlier comments about the quantum total erasure channel. This is depicted in Figures \ref{fig:ps0} and \ref{fig:ps+}. 

A very different situation occurs for $\rf = \ketbra{1}$. In this case only $\ketbra{1}$ is allowed as an output. Thus, the Petz sends $\ketbra{1}$ to itself while all other states are retrodicted in (complicated but logically consistent) ways dependent on channel's forward transitions, reflected in FIG. \ref{fig:ps1}. 

To explain this more symmetrically: in the former two scenarios, all outputs are compatible with the absolute conviction (as enforced by state purity) given to the reference state, hence all outputs are retrodicted to it. Meanwhile, in this latter case, \textit{only one pure output} (which just so happens to be the same as the reference) is compatible with the pure reference state.  Hence, all \textit{other states} (beside the expected output) are retrodicted in accordance to the channel without any regard the reference, since the reference already excludes the possibility of such states.  These more complicated Bayesian inversions come together and cumulate into a vertical reflection of the forward channel, as FIG. \ref{fig:ps1} depicts. For an arbitrary $\rf$, we get a classical mixture of all these key effects together. We depict the case where $\rf = \rf(\frac{\pi }{16},\frac{\pi }{5},\frac{\pi }{3})$ in FIG. \ref{fig:psarbi}.  

It should be said the interplay of reference and channel dependencies we have reviewed here is fundamental in classical retrodiction scenarios as well. The Half-SWAP illustrates that these same Bayesian features hold in the quantum regime via the inferential structure of the Petz Map, even when complementarity and entanglement is introduced.

\section{Conclusions}\label{conclude}
By expressing the Petz Recovery map as a decomposition of matrices given by \eqref{main} we have situated quantum Bayesian inference in the same formal language as that of its classical counterpart given by \eqref{bayesstoch-amended}. We have also highlighted what we have found to be the most noteworthy (and interpretation-neutral) similarities and differences between these two theories when it comes to logical inference. 

\ed{Bayesian inference in both theories involve a similar structure (see $\sret$ in Table \ref{tab:compares}). Given that the transpose of a classical channel is also its adjoint, the key difference between $\pCL$ and $\pQM$ lies not in the central matrix $S^{\chn^\dagger}$, but in the right and left matrices $X_{\chn[\rf]}, X_\rf$ that capture the description of the priors. This affirms the fact that what separates quantum theory from classical theory is not so much in its dynamics (which is in many ways conceptually similar), but in the description of states. 

Mathematically, the difference is captured by the form of the structure coefficients $\xi_{ixjy}$. In classical theory, the structure coefficients render the matrices diagonal. By contrast, in a QPR of quantum theory, the structure coefficients introduce weighted products $v^\alpha_x v^\alpha_y$ of every pair of entries of the distribution $v^\alpha$. This is a consequence of the fact that frames are tomographically complete -- ultimately, a signature of complementarity.}

After these key results we illustrated some examples of quantum Bayesian reasoning using transition graphs, which offer some visual intuitions about how the Petz produces inferences in a quantum regime. Finally, we point to two possible directions to enlarge the perspective open by this work. One may repeat the current study for alternative proposals of quantum Bayes rule that are not equivalent to the Petz map \cite{PF22}. Also, having chosen (quasi)-stochastic processes as the common language for the comparison, we have left out representations of quantum theory that have a different structure and use complex numbers: notably, the Kirkwood-Dirac representation, which has been recently shown to be related to metrological advantages \cite{arv2020}.


\section*{Acknowledgments}
This research was supported by the National Research Foundation and the Ministry of Education, Singapore, under the Research Centres of Excellence programme (till 6 December 2022); and by the National Research Foundation, Singapore, and A$^*$Star under the CQT Bridging Grant (from 7 December 2022 onwards). We also thank Arthur Parzygnat, Jacopo Surace, Zaw Lin Htoo and Eugene Koh for helpful discussions.
\vspace{1em}

\appendix
\section{$S^\chn$ in terms of $\xi_{pqrs}$}\label{WoottDeriv}
\ed{We recall in reference to Table \ref{tab:QHM} that
\begin{equation}\label{chnqpr}
    S^\chn_{ij} = \tr{F_{i} \chn[G_j]}.
\end{equation}
Braasch and Wootters observed (see \cite{braasch2020transition},  Eq.~(19)), in the context of DW representation, that one can write an output state $\rho' = \chn[\rho]$ of a channel $\chn[\bullet] = \sum_l \kappa_l 
\bullet \kappa_l^\dag$ in the following QPR expression:
\begin{equation}
    v^{\rho'}_{i} = \frac{1}{d} \sum_{xyj} \tr{G_{i} G_x G_j G_y} \mathcal{K}_{xy} v^\rho_j,
\end{equation}
where $\mathcal{K}_{xy} = \sum_l K^{(l)}_x K^{(l)*}_y$ and $K^{(l)}_x = \frac{1}{d} \tr{G_x\kappa_l}$. Hence,
\begin{equation}
    v^{\rho'}_{i} = \frac{1}{d^3} \sum_{xylj} \tr{G_x\kappa_l}\tr{G_y\kappa_l^\dag} \tr{G_{i} G_x G_j G_y} v^\rho_j.
\end{equation}
Since $v^{\rho'} = S^\chn v^\rho$, this implies:
\begin{equation}\label{DWchnend}
    (S^\chn_\textbf{DW})_{ij} = \frac{1}{d^3} \sum_{xyl} \tr{G_x\kappa_l}\tr{G_y\kappa_l^\dag} \tr{G_{i} G_x G_j G_y}.
\end{equation}
We build upon this observation by deriving directly from \eqref{chnqpr} an expression akin to \eqref{DWchnend}, but valid for any QPR:
\begin{align}
    S^\chn_{ij} &= \tr{F_{i} \chn[G_j]} \\
    &= \sum_l \tr{F_{i} \kappa_l G_j \kappa_l^\dag} \\
    &= \sum_{xl} \tr{F_x \kappa_l} \tr{G_x G_j \kappa_l^\dag F_{i}} \\
    &= \sum_{xyl} \tr{F_x \kappa_l} \tr{F_y \kappa_l^\dag} \tr{F_{i} G_x G_j G_y} \\ 
    &= \sum_{xyl} \tr{F_x \kappa_l} \tr{F_y \kappa_l^\dag} \xi_{i x j y} \label{ChnComp}
\end{align}
Recalling \eqref{DWDFrame}, we recover \eqref{DWchnend} for DW representations. In \eqref{ChnComp}, we have an expression much like \eqref{closedform}, now catered for general CP maps. We may compare these expressions more instructively, by writing \eqref{closedform} as
\begin{equation}
    S^{\mathcal{M}_\alpha}_{ij}=(M_\alpha)_{ij} = \sum_{xy} \tr{F_x \alpha} \tr{F_y \alpha} \xi_{i x j y}.
\end{equation}
This comparison gives us a sense of what the structure coefficients embed into objects in which they reside. They structurally encode the choice of frame and representation into the entries of these maps, whether they are CPTP or not. 
}

\section{$M_{\alpha^r} = M_{\alpha}^r$ for all $r \in \mathbb Q$}\label{powertopower}

\ed{From the concatenation of maps $S^{\chn} S^{\mathcal{F}} = S^{\chn \circ \mathcal{F}}$ [Eq.~\eqref{concat}] and the definition \eqref{defM}, it is immediate that $M_{\alpha^r} = M_{\alpha}^r$ holds for $r\in\mathbb{Z^+}$. Now we proceed to prove that it is valid for $r\in\mathbb{Q}$.

We recall the entry-wise definition \eqref{m-iso}}
\begin{equation}
    (M_{\alpha_r})_{ij} = \tr{F_i \alpha^r G_j \alpha^r}.
\end{equation}

Firstly, we prove that
\begin{eqnarray*}
    (M_{\alpha^{1\!/2}}M_{\alpha^{1\!/2}})_{ij} &=& \sum_k (M_{\alpha^{1\!/2}})_{ik}(M_{\alpha^{1\!/2}})_{kj} \\ 
    &=& \sum_k \tr{F_i \sqrt{\alpha} G_k \sqrt{\alpha}}\tr{F_k \sqrt{\alpha} G_j \sqrt{\alpha}}\\
    &=& \sum_k \tr{G_k \sqrt{\alpha} F_i \sqrt{\alpha} }\tr{F_k \sqrt{\alpha} G_j \sqrt{\alpha}}\\
    &=& \tr{ \sqrt{\alpha} F_i \sqrt{\alpha} \sqrt{\alpha} G_j \sqrt{\alpha}}\\
    &=& \tr{ F_i \alpha G_j \alpha} = (M_{\alpha})_{ij}.
\end{eqnarray*}
Crucially, in the fourth equality we use the property \eqref{sumtraceproperty} in QPR. Hence, 
\begin{equation*}
    M_{\alpha^{1\!/2}} = M_\alpha^{1\!/2}.
\end{equation*}
\ed{By reiterating this (i.e. sending $\alpha \to \sqrt{\alpha}$), we obtain
\begin{equation*}
    \forall n \in \mathbb{Z^+}: \;  M_{\alpha^{1\!/2n}} = M_\alpha^\frac{1}{2n},
\end{equation*}
Using \eqref{concat} for $N \in \mathbb{Z^+}$, we have,
\begin{eqnarray*}
    M_{\alpha^{N/2n}} &=& S^{\mathcal{M}_{\alpha^{N/2n}}} \\
    &=& S^{\overbrace{\mathcal{M}_{\alpha^{1/2n}}\circ \mathcal{M}_{\alpha^{1/2n}}\circ \dots \circ \mathcal{M}_{\alpha^{1/2n}}}^{N \text{ times}}} \\
    &=& \prod_{N} M_{\alpha^{1/2n}}
    = \prod_{N} M_{\alpha}^{1/2n}.
\end{eqnarray*}
Since any positive rational number can be written as $q=N/2n$, we have proved that
\begin{equation} \label{rooteq}
    M_{\alpha^{q}}= M_{\alpha}^{q}\,,\;\;q\in\mathbb{Q^+}.
\end{equation}
Secondly, we note that 
\begin{eqnarray*}
    (M_{\alpha} M_{\alpha^{-1}})_{ij} &=& \sum_k \tr{F_i \alpha G_k \alpha}\tr{F_k \alpha^{-1} G_j \alpha^{-1}} \\ 
    &=& \sum_k \tr{G_k \alpha F_i \alpha }\tr{F_k \alpha^{-1} G_j \alpha^{-1}} \\
    &=& \tr{\alpha F_i \alpha \alpha^{-1} G_j \alpha^{-1}} \\
    &=& \tr{F_i G_j} = \delta_{ij} = \one_{ij}.
\end{eqnarray*}
Hence, $M_{\alpha^{-1}} = M_{\alpha}^{-1}$. Repeating this, we can easily see that for any $N \in \mathbb{Z}^+$:
\begin{equation} \label{negeq}
    M_{\alpha^{-N}} = M_{\alpha}^{-N}.
\end{equation}
Finally, taking from \eqref{concat}, \eqref{rooteq} and \eqref{negeq}, we find that:
\begin{eqnarray*}
    M_{\alpha^{q-N}} &=& S^{\mathcal{M}_{\alpha^{q-N}}} = S^{\mathcal{M}_{\alpha^{q}}\circ \mathcal{M}_{\alpha^{-N}}} \\
    &=& M_{\alpha^{q}} M_{\alpha^{-N}} = M_{\alpha}^{q} M_{\alpha}^{-N}\\ 
    &=& M_{\alpha}^{q-N}.
\end{eqnarray*}
Since any rational number can be written as a positive rational number minus a positive integer, we have proved that $M_{\alpha^r} = M_{\alpha}^r$ for any $r \in \mathbb Q$.}


\section{$S^{\chn^\dagger}$ in terms of $(S^{\chn})^\text{T}$ for QPRs}\label{adjoints}
We derive the QPR expressions for $S^{\chn^\dagger}$ for some CP map $\chn[\bullet] = \sum_l \kappa_l \bullet  \kappa_l^\dagger$. For NQPRs, we find easily that: 
\begin{eqnarray*}
    (S^{\chn^\dagger}_\textbf{NQ})_{ij} &=& \tr{F_i \chn^\dagger[G_j]} =  \tr{F_i \sum_l \kappa_l^\dagger G_j \kappa_l} \\ &=& \sum_l \tr{ G_j \kappa_l F_i  \kappa_l^\dagger} = \sum_l \tr{ F_j \kappa_l G_i  \kappa_l^\dagger} \\ &=&  \tr{ F_j \sum_l \kappa_l G_i  \kappa_l^\dagger} = \tr{ F_j \chn[G_i]} = S^\chn_{ji}.
\end{eqnarray*}
Thus, for NQPRs $S^{\chn^\dagger}_\textbf{NQ} = (S^\chn)^\text{T}$. For SIC-POVM representations, we have a more complicated expression. We first use \eqref{SPconvert},
\begin{eqnarray*}
    (S^{\chn^\dagger}_\textbf{SP})_{ij} &=& \tr{F_i \chn^\dagger[G_j]} \\
    &=& \tr{\frac{G_i+\one}{d(d+1)} \, \chn^\dagger\left[ d(d+1) F_j - \one \right]}. 
\end{eqnarray*}
By expanding the terms and noting the unitality of every adjoint map (i.e. $\chn^\dagger [\one] = \one$), we arrive at the expression: 
\begin{eqnarray*}
    (S^{\chn^\dagger}_\textbf{SP})_{ij} &=& \tr{F_i \sum_l \kappa_l^\dagger G_j \kappa_l} + \tr{\chn^\dagger[F_{j}]} - \tr{F_i} \\
    &=& S^\chn_{ji} + \tr{\chn^\dagger[F_{j}]} - \frac{1}{d}.
\end{eqnarray*}
By taking note of the relations found in Table \ref{tab:QHM}, we may write $\tr{\chn^\dagger[F_{j}]} = \sum_l \tr{ \kappa_l^\dagger F_j \kappa_l} = \sum_l \tr{  F_j \kappa_l \one \kappa_l^\dagger} = \tr{  F_j \chn[\one]} = \frac{1}{d}(S^\chn v^1)_j = \frac{1}{d}\sum_a \chn(j|a)$. Hence, we can write the total expression of each entry for SIC-POVM representation as: 
\begin{equation}
    (S^{\chn^\dagger}_\textbf{SP})_{ij} = S^\chn_{ji} + \frac{1}{d}\left(\sum_a \chn(j|a)-1 \right).
\end{equation}
This can be written, on the matrix level, as \eqref{SPadjoint}.

\section{$(S^{\chn})^\text{T}$ as $S^{\chn^{\dagger}}$ for Classical} \label{classicaladjoint}
In the previous section we proved that for quantum channels (expressed in QPRs), we can express the adjoint channel in terms of the tranpose of the channel. Here, we prove the opposite relation for classical channels: that the transpose of a classical channel is the adjoint of that channel. Namely, the transpose map is the map for which \eqref{eq:adj} is fulfilled in the case of classical scenarios. Noting first the commutative diagram found in FIG. \ref{tracecommdiag} (which invokes the relationships found in Table \ref{tab:QHM}), we see how \eqref{eq:adj} is fulfilled by a map for which 
\begin{equation}\label{adjtotp}
    (S^\mathcal{E} v^\rho) \cdot \bar{v}^\sigma = (S^\mathcal{E^\dag} v^\sigma) \cdot \bar{v}^\rho,
\end{equation}
for all $\rho$ and $\sigma$.
\begin{figure}[t!]
    \centering
\begin{tikzcd}[sep=1.6em]
	 {v^\rho \cdot \bar{v}^\sigma} && {\text{Tr}[\rho\,\sigma]} \\
	{(S^\mathcal{E} v^\rho) \cdot \bar{v}^\sigma}&& {\text{Tr}\left[\mathcal{E}[\rho]\sigma\right]} \\
	{(S^\mathcal{E^\dag} v^\sigma) \cdot \bar{v}^\rho} && {\text{Tr}\left[\mathcal{E}^\dag[\sigma]\rho\right]} 
	\arrow[leftrightarrow, from=1-1, to=1-3]
	\arrow[ from=1-1, to=2-1]
	\arrow[ from=2-1, to=3-1]
	\arrow[leftrightarrow,  from=2-1, to=2-3]
	\arrow[leftrightarrow,  from=3-1, to=3-3]
	\arrow[from=1-3, to=2-3]
	\arrow[from=2-3, to=3-3]
        \arrow["\forall\rho\sigma", from=2-3, to=3-3]
        \arrow["\forall\rho\sigma", from=2-1, to=3-1]
\end{tikzcd}
    \caption{Relations within and functors between formalisms pertaining the adjoint map, illustrated commutatively.}
    \label{tracecommdiag}
\end{figure}
With this, we expand the LHS of \eqref{adjtotp}:
\begin{eqnarray} 
    (S^\mathcal{E} v^\rho) \cdot \bar{v}^\sigma  &=& \sum_y (S^\mathcal{E} v^\rho)_y \bar{v}^\sigma_y \\
    &=& \sum_{xy} S^\mathcal{E}_{yx} v^\rho_x \bar{v}^\sigma_y .\label{adjcl}
\end{eqnarray}
Next we expand the following, in order to check if the transpose qualifies as the adjoint:
\begin{eqnarray}
    ((S^\mathcal{E})^\text{T} v^\sigma) \cdot \bar{v}^\rho &=& \sum_{xy} (S^\mathcal{E})_{yx}^{\text{T}} v^\sigma_x \bar{v}^\rho_y \\
    &=& \sum_{xy} S^\mathcal{E}_{xy} v^\sigma_x \bar{v}^\rho_y \\ &=& \sum_{xy} S^\mathcal{E}_{yx} \bar{v}^\rho_x v^\sigma_y.  \label{transcl}
\end{eqnarray}
Now for classical scenarios the trace of two states, if treated like quantum states in Hilbert space, would simply be the inner product of its density spectra: ${v^\rho \cdot {v}^\sigma} = {\text{Tr}[\rho\,\sigma]}$. This is because the states, being classical distributions, would be diagonalized in the same way. Thus we could have replaced $\bar{v}^\rho$ with $v^\rho$ in all the above calculations and in FIG. \ref{tracecommdiag}. The reason why we have written $\bar{v}^\rho$ as opposed to ${v}^\rho$ is to simply highlight that while indeed \eqref{adjcl} is identical to \eqref{transcl} for classical scenarios because $\bar{v}^\rho = v^\rho$ there (and NQPR for that matter since $\bar{v}^\rho = c\, v^\rho$), the same does \textit{not} hold for SIC-POVM. The transpose qualifies as an adjoint for both NQPR and classical channels, but not for SIC-POVM. Hence, the relation proved for classical states and channels does not contradict the ones proved in the previous section for QPRs. 

\section{Properties of $M_{\alpha}$} \label{propertiesofMalpha}
Here we note some interesting properties of $M_{\alpha}$. Namely that it is a matrix with all real entries and non-negative eigenvalues that sum to 1. 

\subsection{Real Entries}
It can be shown that all the entries of $M_\alpha$ are real: $(M_{\alpha})_{ij} = \tr{F_i \alpha G_j \alpha} \in \mathbb R$. A proof was given in the main text, valid for any $M_{\alpha^r}$; we repeat it here for completeness.

We first note that the anticommutator for any two Hermitian operators $A$ and $B$ is always also Hermitian: $\{A,B\}^\dagger=\{A,B\}$; while the trace of the commutator of any two operators is always zero (in finite dimension) due to cyclicity: $\text{Tr}\big[[A,B]\big]=\tr{AB}-\tr{BA}=0$.
Hence 
\begin{equation}\label{realofprodherm}
    \tr{AB} = \tr{\frac{\{A,B\}}{2}+\frac{[A,B]}{2}}= \frac{1}{2}\tr{\{A,B\}} \in \mathbb{R}
\end{equation}
Noting that $\sqrt{\alpha} F_i \sqrt{\alpha}$ and $\sqrt{\alpha} G_j \sqrt{\alpha}$ are both Hermitian (frame and dual operators are always Hermitian, and $\alpha$ is a density operator in Hilbert Space), we apply \eqref{realofprodherm} to $(M_{\alpha})_{ij}$. The entries of $M_\alpha$ are thus proven to be always real.

\subsection{Positive Semi-Definiteness}
For any NQPR, we can always write 
\begin{equation*}
    (M_{\alpha})_{ij} = c \, \tr{\sqrt{\alpha} F_i \sqrt{\alpha} \, \big(\sqrt{\alpha} F_j \sqrt{\alpha} \big)^\dagger} .
\end{equation*} 
Hence, $M_{\alpha}$ is a Gram matrix with some positive factor $c$. Thus it is positive semi-definite. For SIC-POVM, we expand $(M_{\alpha})_{ij}$ via \eqref{SPconvert}, arriving at: 
\begin{eqnarray*}
     (M_{\alpha})_{ij} = \frac{1}{d(d+1)} \Big(  \tr{\sqrt{\alpha} 
        G_{i}  \sqrt{\alpha} \, \big(\sqrt{\alpha} G_j \sqrt{\alpha} \big)^\dagger} \\ 
        + \tr{G_j \alpha^2} \Big).
\end{eqnarray*}
The first term, as with the NQPR case, corresponds to a Gram Matrix, which is positive semi-definite. One can then note that the second term corresponds to a matrix $J_\alpha$ (i.e. $(J_\alpha)_{ij} = \tr{G_j \alpha^2}$) with duplicate rows (every $j$-th column with filled with identical entries. This simply implies that the only non-zero eigenvalue would be the sum of the entries of any given row. Which just means: $\text{eig}[ J_\alpha ] = \{\sum_j \tr{G_j \alpha^2} , 0\} = \{\tr{\alpha^2} , 0\} \geq 0$. So $M_\alpha$ is the sum of two positive semi-definitive matrices and thus we may conclude $M_{\alpha} \geq 0$ for SIC-POVM as well. 
\subsection{Unit Trace}
The trace of $M_\alpha$ is given by
\begin{eqnarray*}
    \tr{M_\alpha} = \sum_{i} (M_{\alpha})_{ii} = \text{Tr}\Big[\underbrace{\sum_i F_i \alpha G_i}_{\one} \, \alpha \Big] = 1 .
\end{eqnarray*}
To prove the relation invoked for the final equality we will use the previously found result in \cite{renes2004symmetric}. Consider the superoperator
\begin{equation}
    \Lambda[\bullet] = \sum_{i=1}^{d^2}\Pi_i \bullet \Pi_i ,
\end{equation}
it can be shown that 
\begin{equation}
    \Lambda[\Pi_i] = \frac{d}{d+1}(\Pi_i + \one).
\end{equation}
Since the set $\{\Pi_i\}$ forms a basis, we can express the superoperator as
\begin{equation}
    \Lambda = \frac{d}{d+1}(\mathcal{I} + \one),
\end{equation}
where $\mathcal{I}[A] = \tr{A}\one$. Using this we can easily show that for SIC-POVM representation we have
\begin{eqnarray}
    \sum_{i}F_i\alpha G_i &=& \frac{d+1}{d}\sum_i \Pi_i \alpha \Pi_i - \frac{1}{d}\alpha \sum_i \Pi_i \nonumber \\
    &=& \one\, .
\end{eqnarray}
For discrete Wigner representation, Zhu \cite{zhu2016quasiprobability} showed that the dual frame can always be expressed as such:
\begin{equation}
    G_i = -\sqrt{d+1}\Pi_i + \left(\frac{1+\sqrt{d+1}}{d}\right)\one .
\end{equation}
Thus, it can also be easily shown that $\sum_{i}F_i\alpha G_i = \one$ in this representation.

\section{General Examples for $\pQM \neq \pCL$} \label{Q-is-not-C}
As discussed in Section \ref{PetzforHSWAP}, it is the case that $\hat{\chn}_{\rf}[\rho]=\hat{\chn}_{+}[\rho]=\ketbra{+}$ for all $\rho$ when $\chn[\bullet]=\TrB[\pswap \bullet \otimes \ketbra{1} \, \pswap^\dagger]$ and $\rf=\ketbra{+}$. Yet we can easily find that, for the canonical state representations for DW and SIC-POVM, we have:
\begin{eqnarray*}
S^{\tilde{\chn}_{+}}_{\textbf{DW}} &=& 
\begin{psmallmatrix}
1 & \frac{1}{7} \left(3-\sqrt{2}\right) & 1 & \frac{1}{7} \left(\sqrt{2}+3\right) \\
 0 & \frac{1}{7} \left(\sqrt{2}+4\right) & 0 & \frac{1}{7} \left(4-\sqrt{2}\right) \\
 0 & 0 & 0 & 0 \\
 0 & 0 & 0 & 0 \\ 
\end{psmallmatrix} \\
&\neq& \frac{1}{2}
\begin{psmallmatrix}
 1 & 1 & 1 & 1 \\
 1 & 1 & 1 & 1 \\
 0 & 0 & 0 & 0 \\
 0 & 0 & 0 & 0 \\
\end{psmallmatrix} = S^{{\hat{\chn}_{+}}}_{\textbf{DW}}
\end{eqnarray*}
Likewise, 
\begin{eqnarray*}
S^{\tilde{\chn}_{+}}_{\textbf{SP}} &=& 
\begin{psmallmatrix}
 0.925 & 0.183 & -0.264 & 0.353 \\
 0.0744 & 0.744 & 0.275 & 0.168 \\
 -0.0191 & 0.0491 & 0.915 & 0.0947 \\
 0.0199 & 0.0233 & 0.0737 & 0.384 \\
\end{psmallmatrix} \\
&\neq& \frac{1}{12}
\begin{psmallmatrix}
 \sqrt{3}+3 & \sqrt{3}+3 & \sqrt{3}+3 & \sqrt{3}+3 \\
 \sqrt{3}+3 & \sqrt{3}+3 & \sqrt{3}+3 & \sqrt{3}+3 \\
 3-\sqrt{3} & 3-\sqrt{3} & 3-\sqrt{3} & 3-\sqrt{3} \\
 3-\sqrt{3} & 3-\sqrt{3} & 3-\sqrt{3} & 3-\sqrt{3} \\
\end{psmallmatrix} = S^{{\hat{\chn}_{+}}}_{\textbf{SP}}
\end{eqnarray*}
Indeed, for some channels one can find states for which the post-measurement probabilities violate acceptable bounds. This means $\pCL$ fails to represent a generally valid quantum transformation. For instance, for a unitary transformation $\mathcal{U}[\bullet]= U\bullet U^\dagger$ where $U = \frac{i}{2}\begin{psmallmatrix}
 \sqrt{3} & -1 \\
 1 & \sqrt{3} \\
\end{psmallmatrix}$ We find that 
\begin{eqnarray*}
    (S^{\tilde{\mathcal{U}}_{+}}_{\textbf{DW}} v^{+})\cdot \bar{v}^{0} &=& \frac{1}{2}(1+\sqrt{3}) > 1 \\
    (S^{\tilde{\mathcal{U}}_{+}}_{\textbf{SP}} v^{0})\cdot \bar{v}^{+} &=& \frac{1}{13}(2-5 \sqrt{3}) < 0.
\end{eqnarray*}
$S^{\petz} \neq \pCL$ is thus easily shown.

\section{$\pQM \neq \pCL$ even when $\chn[\rf], \rf, \rho, E_m$ all commute} \label{CounterExamplesCQ}
\ed{Some may expect that quantum retrodiction, under QPR formalism and using its mathematical equipment, would go to classical retrodiction once all states and transformations share the same eigenbasis. Put differently, there may be an expectation that, for every choice of QPR and if $\chn[\rf], \rf, \rho, E_m$ commute, $(\pQM v^{\rho}) \cdot \bar{v}^m = (\pCL v^{\rho}) \cdot \bar{v}^m$. 

However, it turns out that this is not the case. We simply list two examples where this does not obtain. For the canonical choice of DW representation,
\begin{align*}
U &= \frac{1}{4} 
\begin{psmallmatrix}
 \sqrt{2}+2 & 1-i & -1+i & i \left(\sqrt{2}-2\right) \\
 -1-i & \sqrt{2}+2 & 2-\sqrt{2} & 1-i \\
 1+i & 2-\sqrt{2} & \sqrt{2}+2 & -1+i \\
 -i \left(\sqrt{2}-2\right) & -1-i & 1+i & \sqrt{2}+2 \\
\end{psmallmatrix}
 \\
\beta &= \frac{1}{2}\one, \; \rf = \rho = E_m = \ketbra{-\alpha}{-\alpha}
\end{align*}
Where $U$ and $\beta$ define the unitary dilation for $\chn$, $\ket{+\alpha } = \cos(\pi/4) \ket{0} +e^{i \pi/8} \sin(\pi/4) \ket{1}$, $\ket{-\alpha} = e^{-i \pi/8} \sin(\pi/4)  \ket{0} - \cos(\pi/4)\ket{1}$. Essentially this gives a simple noise-inducing channel (for qubits) that preserves the coherence of states in the $\ket{+\alpha}$ and $\ket{-\alpha}$ basis. It can be found that: 
\begin{align*}
    (\pQM v^{\rho})^\text{T} &= \frac{1}{4} (  1-\sqrt{2}, \; 1, \;  1 + \sqrt{2}, \; 1 ) \\ 
    (\pCL v^{\rho})^\text{T} &= \frac{1}{16}(  8-5\sqrt{2}, \; 2 + \sqrt{2}, \; 4 + 3\sqrt{2},\; 2 + \sqrt{2} ) \\
    (\pQM v^{\rho}) & \cdot \bar{v}^m = \frac{1}{2}, \; \; (\pCL v^{\rho}) \cdot \bar{v}^m = \frac{1}{16} (8 - \sqrt{2})
\end{align*}
Similarly, for the canonical choice of frames under SIC-POVM: 
\begin{align*}
U &= \pswap, \;
\rf = \rho = \frac{1}{2}\one,\\ \beta &= \ketbra{0}{0},\; E_m = \ketbra{1}{1}
\end{align*}
Here, we have an amplitude damping channel for the $\ket{0}$, $\ket{1}$ basis. It can be found that: 
\begin{align*}
    (\pQM v^{\rho})^\text{T} &= \frac{1}{36}(  9-\sqrt{3}, \;9+\sqrt{3}, \; 9-\sqrt{3},\; 9+\sqrt{3} ) \\ 
    (\pCL v^{\rho})^\text{T} &= \frac{1}{44} (  11-\sqrt{3}, \; 11+\sqrt{3}, \;  11-\sqrt{3}, \; 11+\sqrt{3} ) \\
    (\pQM v^{\rho}) & \cdot \bar{v}^m = \frac{7}{11}, \; \; (\pCL v^{\rho}) \cdot \bar{v}^m = \frac{2}{3}
\end{align*}
So even when all relevant states and POVMs are diagonal in Hilbert space, the QPR formalism does not make it such that quantum Bayesian inference goes to the classical Bayes rule. This highlights the categorical difference between quantum and classical theory even when they are using similar mathematical equipment. Once one employs tomographically complete frames in ones' QPR, one may not simply re-invoke classical Bayes rule hoping it will simulate the quantum retrodiction, even when all quantum states and transformations commute.

Notice that this does not contradict the known result, mentioned in Section \ref{intro}, that the Petz map \eqref{petz} reduces to classical Bayes rule when all relevant quantum objects are diagonal in the same basis. This is because that result holds in the context of the $d$-dimensional Hilbert space formalism. Meanwhile the above conclusion is made under a QPR, when a valid frame has been affixed to define $d \times d$ real space.} 

\section{Other Transition Graphs}\label{app:graphs}

\begin{figure}[!htb]
\begin{subfigure}{0.45\textwidth}
\centering
         \includegraphics[width=0.93\linewidth]{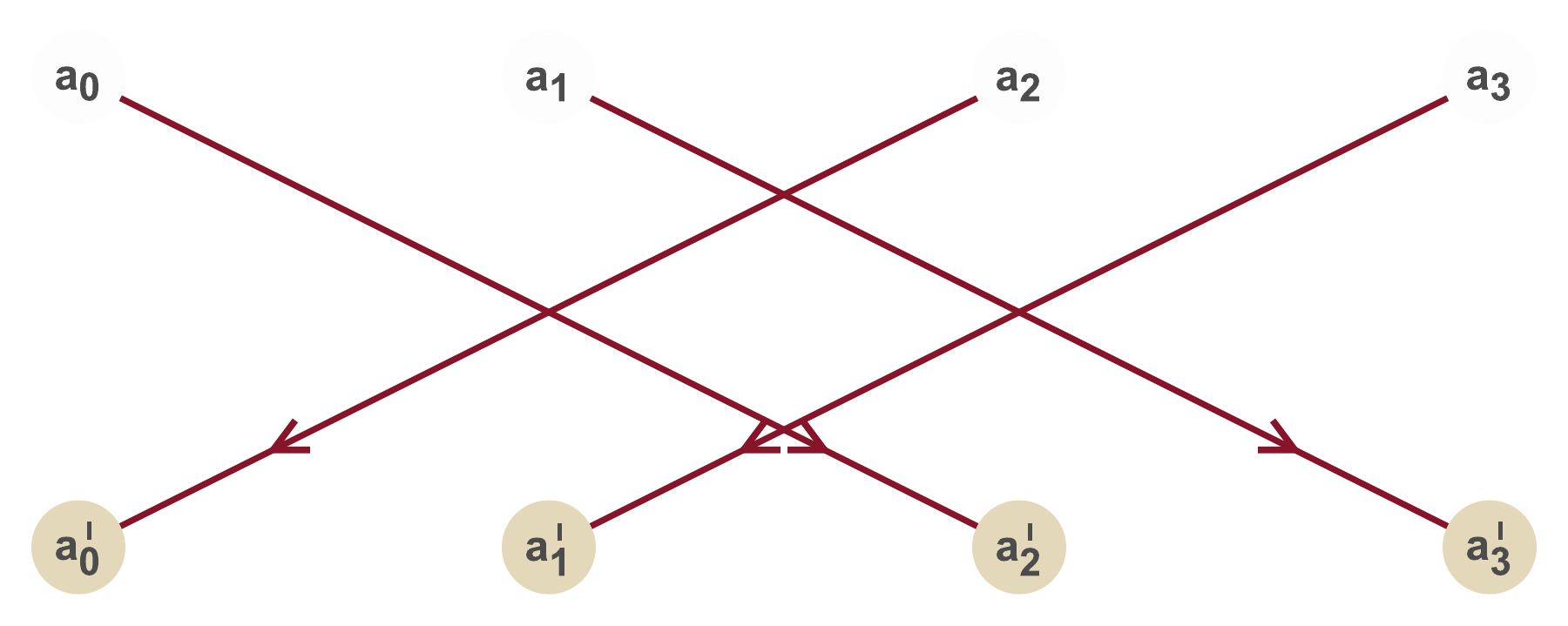}
         \caption{$S^{\chn}$ for $\chn[\bullet]=\sigma_z \bullet\sigma_z$}
\end{subfigure} 
\begin{subfigure}{0.45\textwidth}
\centering
\includegraphics[width=0.93\linewidth]{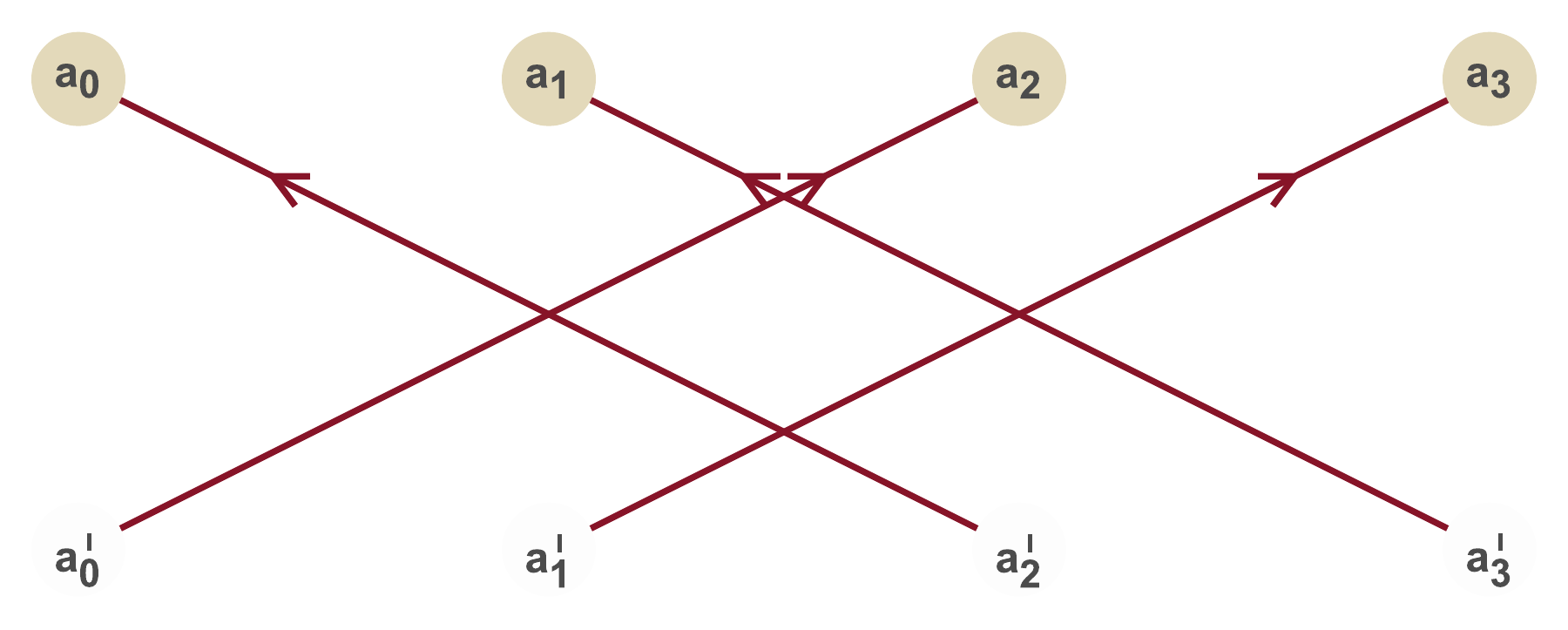}
\caption{$\pSP$ and $\pDW$ for $\chn[\bullet]=\sigma_z \bullet\sigma_z$}
     \end{subfigure}
\begin{subfigure}{0.45\textwidth}
\centering
         \includegraphics[width=0.93\linewidth]{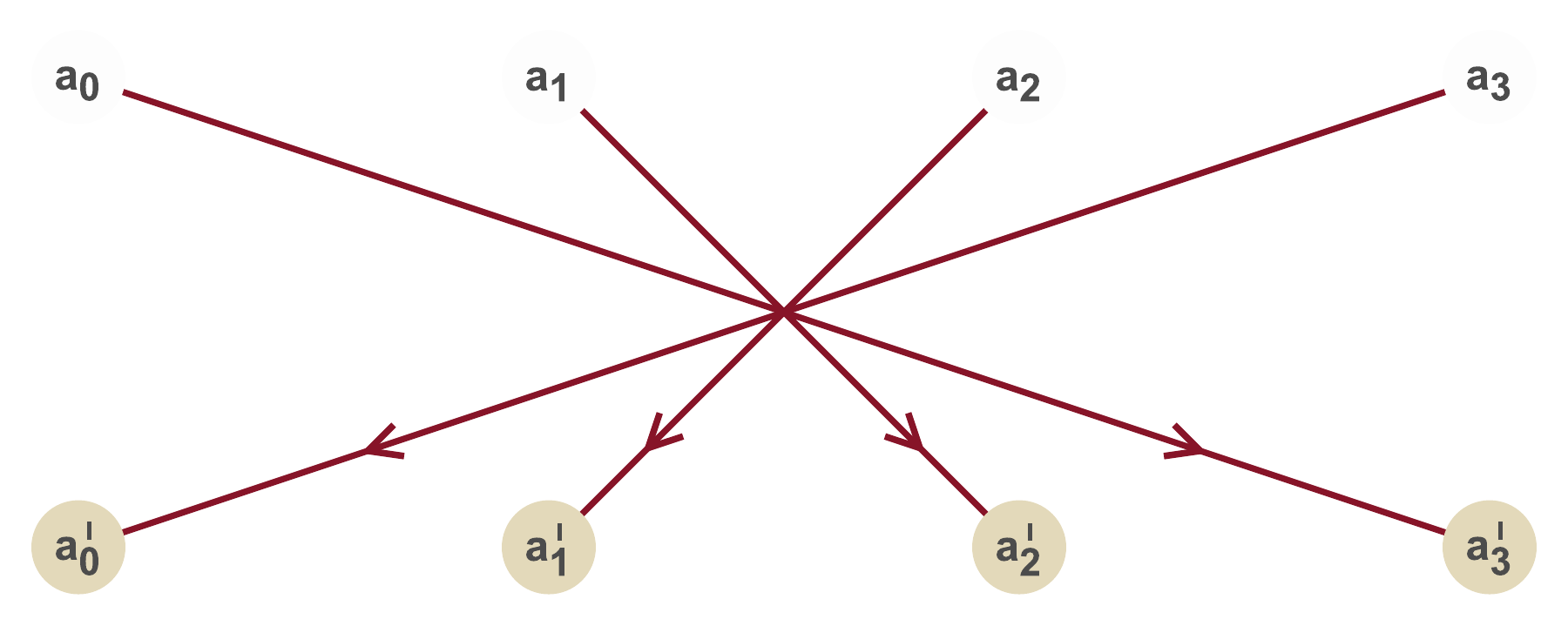}
         \caption{$S^{\chn}$ for $\chn[\bullet]=\sigma_y \bullet\sigma_y$}
\end{subfigure} 
\begin{subfigure}{0.45\textwidth}
\centering
\includegraphics[width=0.93\linewidth]{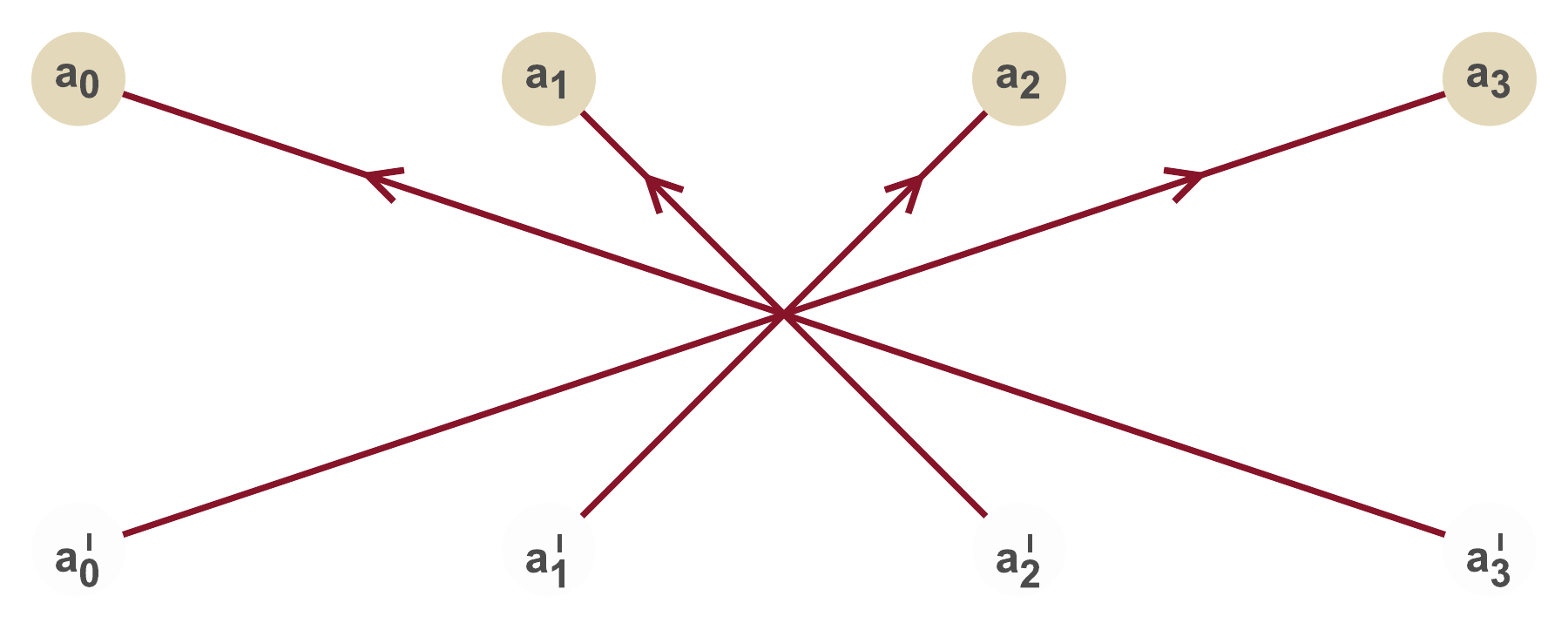}
\caption{$\pSP$ and $\pDW$ for $\chn[\bullet]=\sigma_y \bullet\sigma_y$}
     \end{subfigure} 
\begin{subfigure}{0.45\textwidth}
\centering
         \includegraphics[width=0.93\linewidth]{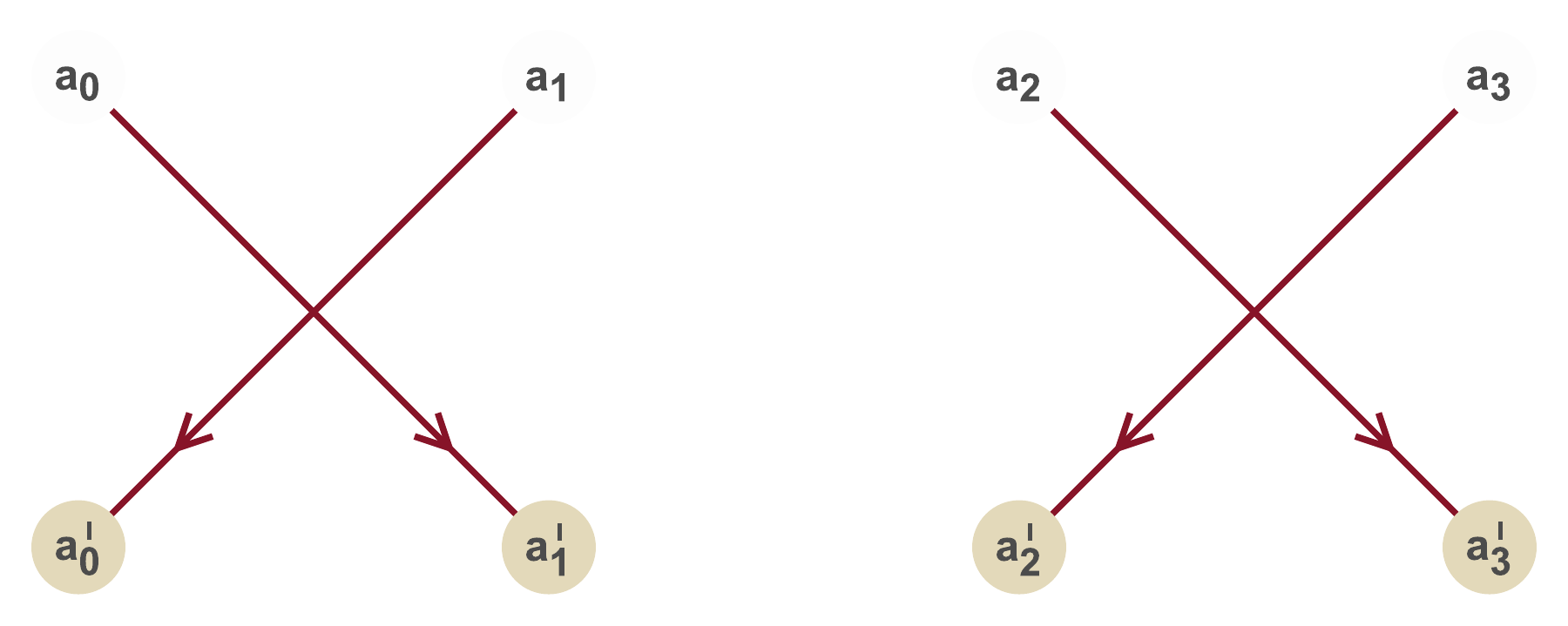}
         \caption{$S^{\chn}$ for $\chn[\bullet]=\sigma_x \bullet\sigma_x$}
\end{subfigure} 
\begin{subfigure}{0.45\textwidth}
\centering
\includegraphics[width=0.93\linewidth]{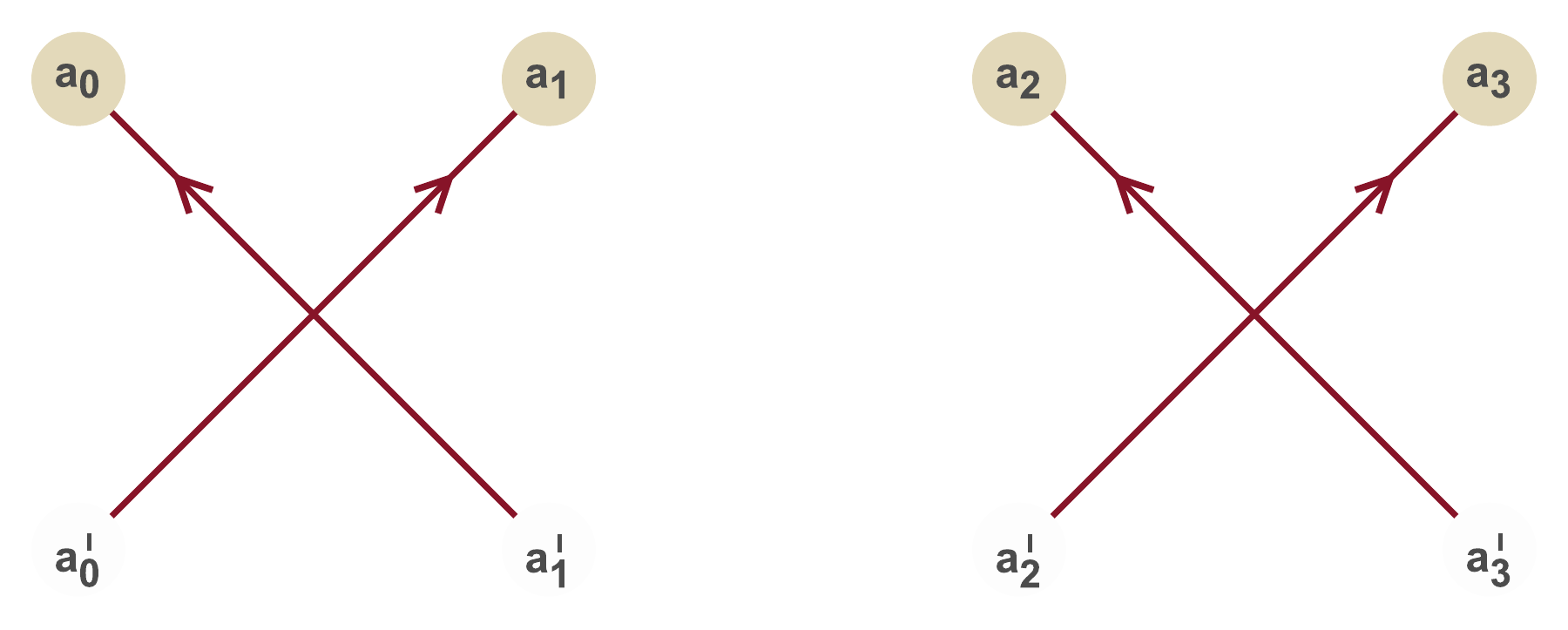}
\caption{$\pSP$ and $\pDW$ for $\chn[\bullet]=\sigma_x \bullet\sigma_x$}
     \end{subfigure} 
     \caption{Transition Graphs for $\sigma_z, \sigma_x$ and $\sigma_y$ and their respective retrodictions.}
     \label{fig:perm}
\end{figure}

In this appendix, we include illustrative cases of $S^\chn$, some respective retrodictions and their transition graphs. In FIG. \ref{fig:perm}, the transition graphs are depicted for very familiar Pauli rotations. It so happens that these unitaries translate to $S^\chn$ that give permutations. This is seen in the bold bijective transition arrows. Like other unitary channels, all retrodictions are reference-prior independent. Transition graphs of such retrodictions are thus always mirror images of the corresponding forward transition graph. That said, most unitaries do not enjoy a permutative structure that exists for these SU(2) rotations. The Hadamard gate for instance defined by the following computationally represented operator and gives the respective quasi-stochastic matrix: 
\begin{eqnarray*}
 U_{\text{H}} \, \hat{=} \, \frac{1}{\sqrt{2}}\begin{pmatrix}
    1 & 1 \\ 1 & -1 
\end{pmatrix}, \quad S^{\mathcal{U}_{\text{H}}} &=& \frac{1}{2} 
\begin{psmallmatrix}
1 & 1 & 1 & -1 \\
1 & -1 & 1 & 1 \\
1 & 1 & -1 & 1 \\
 -1 & 1 & 1 & 1 
\end{psmallmatrix},
\end{eqnarray*}
which is consistent across the canonical choices of the DW and SIC-POVM representations. 

\begin{figure}[b] 
\begin{subfigure}{0.45\textwidth}
\centering
         \includegraphics[width=0.95\linewidth]{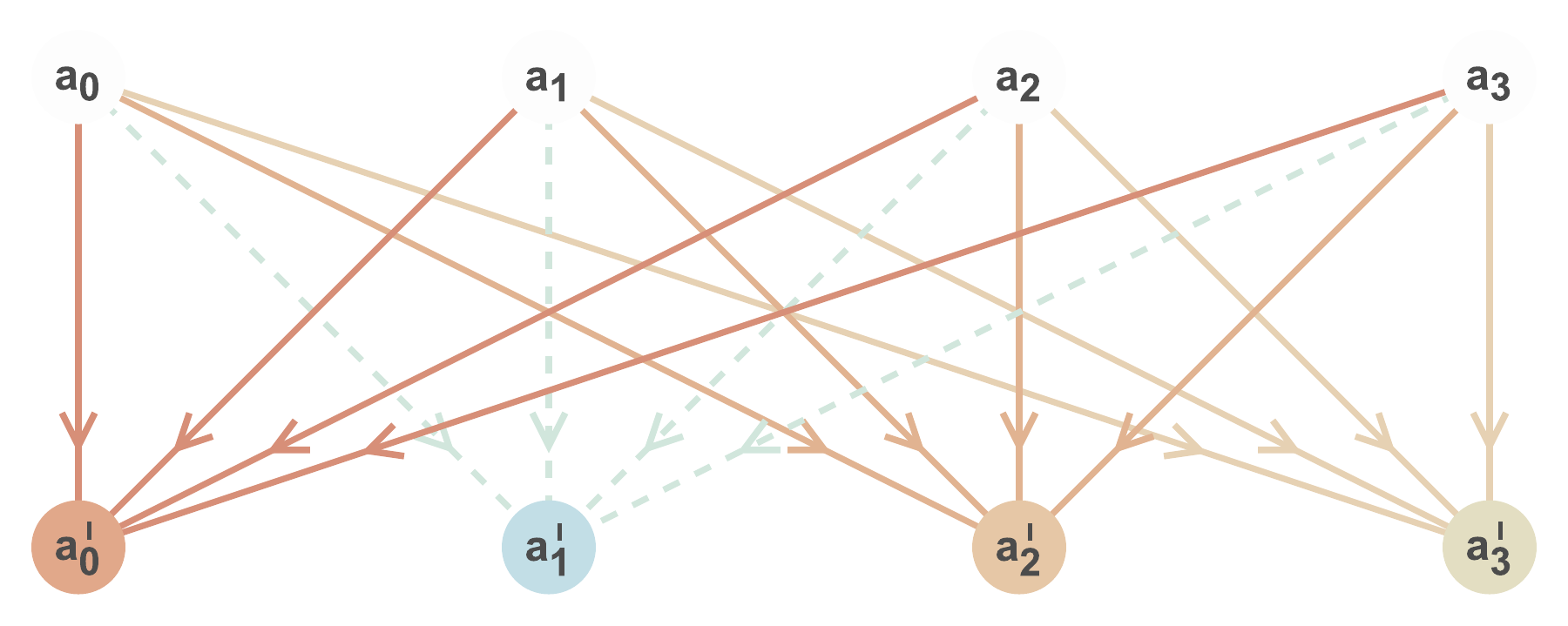}
         \caption{$S^{\chn}_\textbf{DW}$ for $\chn[\bullet]=\TrB[\swap \bullet \otimes \beta(\frac{7 \pi }{16},\frac{3 \pi }{5},\frac{\pi }{6}) \, \swap^\dagger]$}
\end{subfigure} 
\begin{subfigure}{0.45\textwidth}
\centering
         \includegraphics[width=0.93\linewidth]{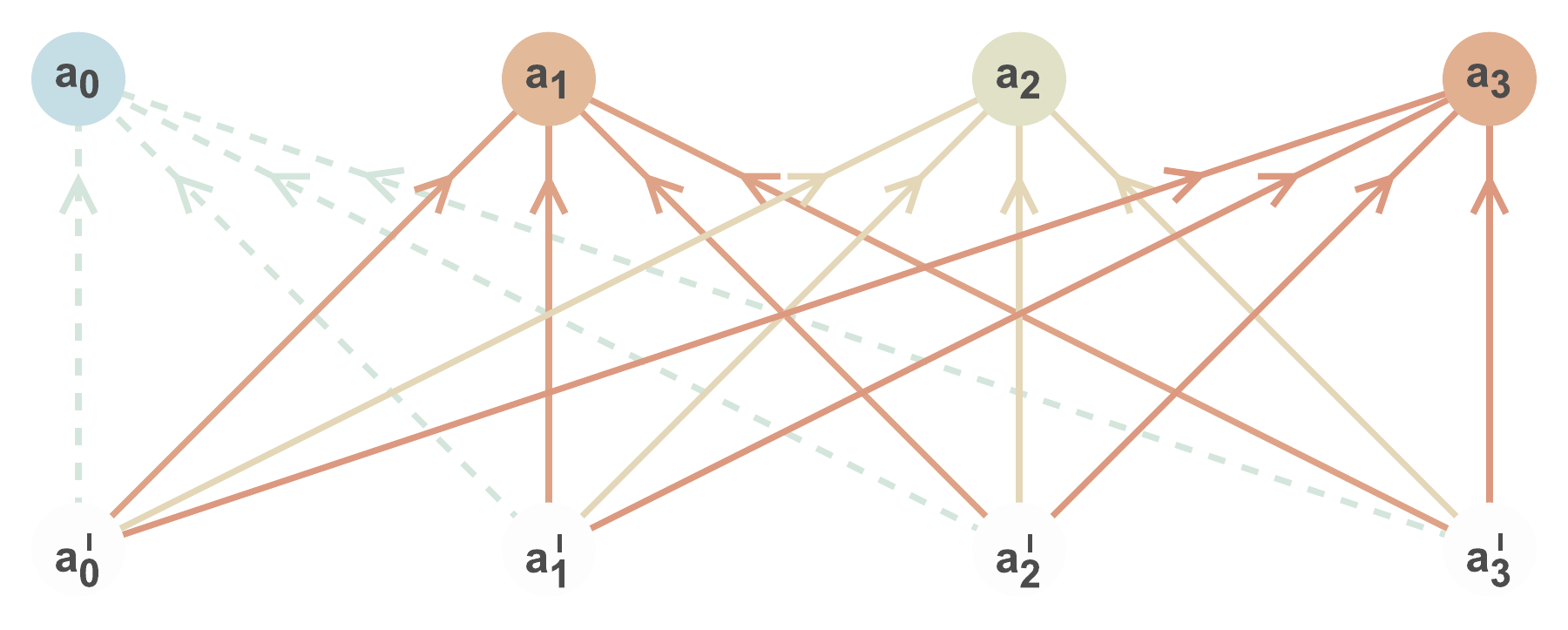}
         \caption{$\pDWW$ for $\swap, \beta(\frac{7 \pi }{16},\frac{3 \pi }{5},\frac{\pi }{6}), \rf(\frac{\pi }{16},\frac{\pi }{5},\frac{\pi }{8}) $}
\end{subfigure} 
     \caption{Transition Graphs for a Quantum Total Erasure channel with arbitrary ancilla $\beta$ and a corresponding retrodiction with reference prior $\rf$.}
     \label{fig:SWAP}
\end{figure} 

\begin{figure}[!htb] 
\begin{subfigure}{0.45\textwidth}
\centering
         \includegraphics[width=0.93\linewidth]{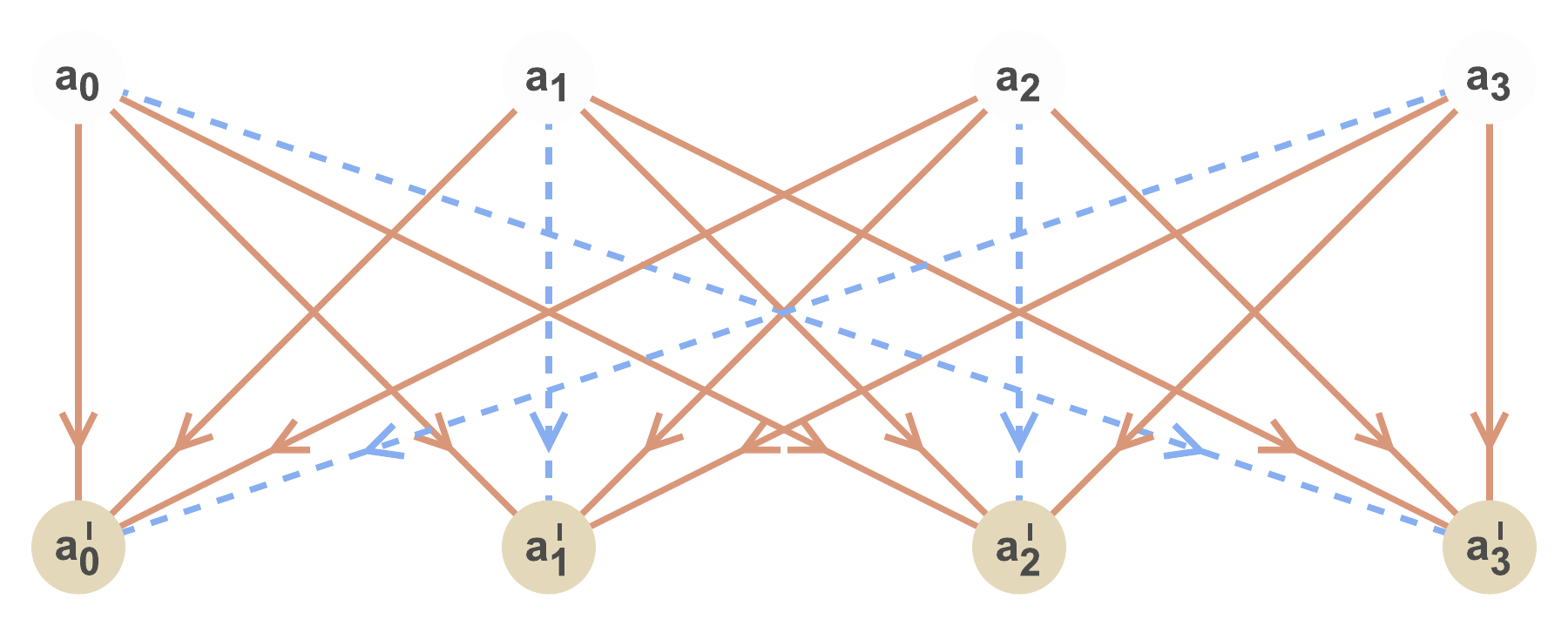}
         \caption{$S^{\mathcal{U}_{\text{H}}}$ for $\mathcal{U}_{\text{H}}[\bullet]=U_{\text{H}} \bullet U_{\text{H}}^\dagger $}
\end{subfigure} 
\begin{subfigure}{0.45\textwidth}
\centering
         \includegraphics[width=0.93\linewidth]{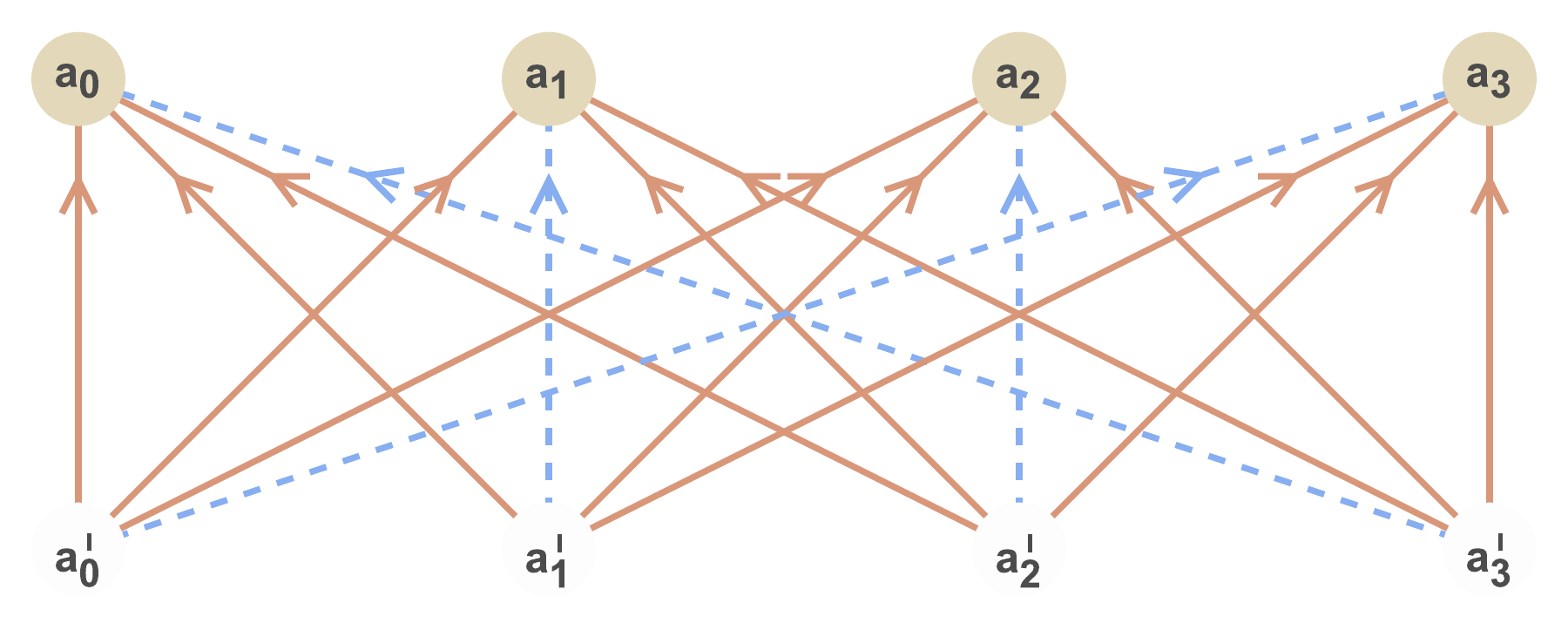}
         \caption{$S^{\hat{\mathcal{U}}_{\text{H}}}$ for $\mathcal{U}_{\text{H}}[\bullet]=U_{\text{H}} \bullet U_{\text{H}}^\dagger $}
\end{subfigure} 
\begin{subfigure}{0.45\textwidth}
\centering
         \includegraphics[width=0.93\linewidth]{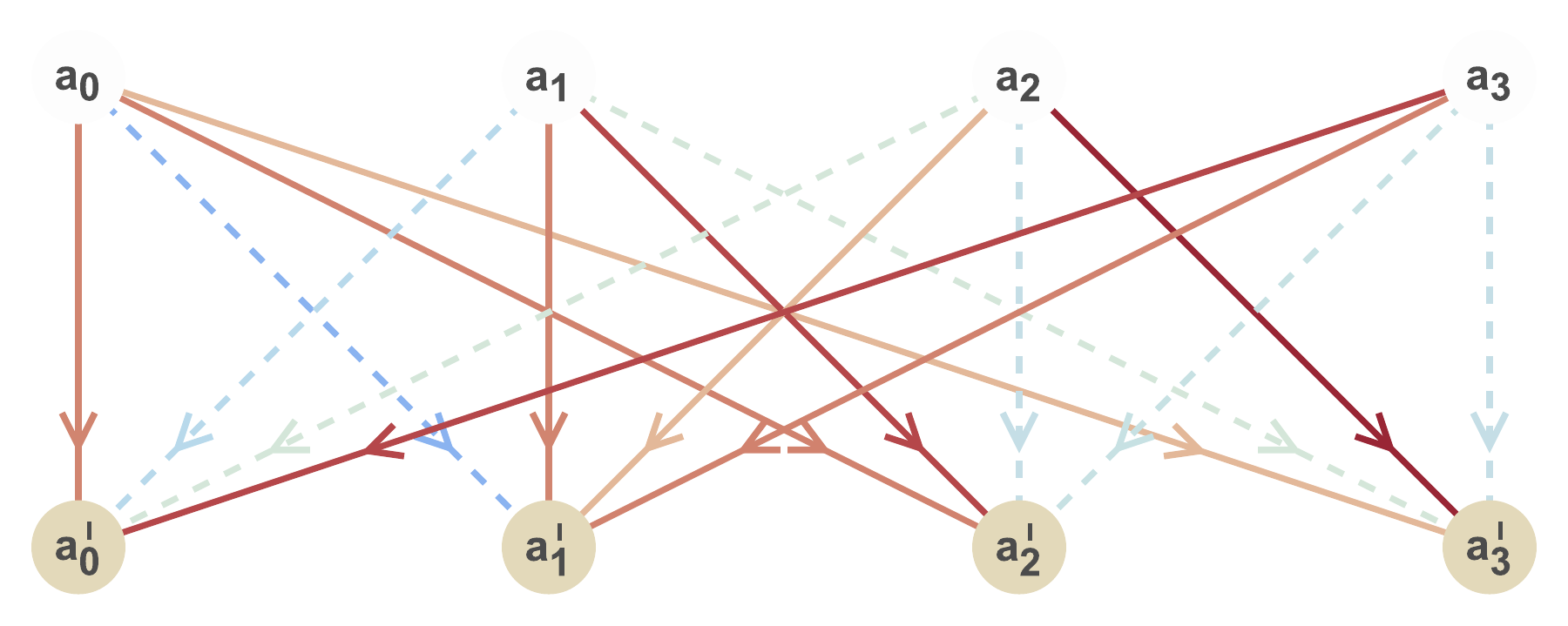}
         \caption{$S^{\mathcal{U}_{\text{eg}}}_\textbf{DW}$ for $\mathcal{U}_{\text{eg}}[\bullet]=U_{\text{eg}} \bullet U_{\text{eg}}^\dagger $}
\end{subfigure} 
\begin{subfigure}{0.45\textwidth}
\centering
         \includegraphics[width=0.93\linewidth]{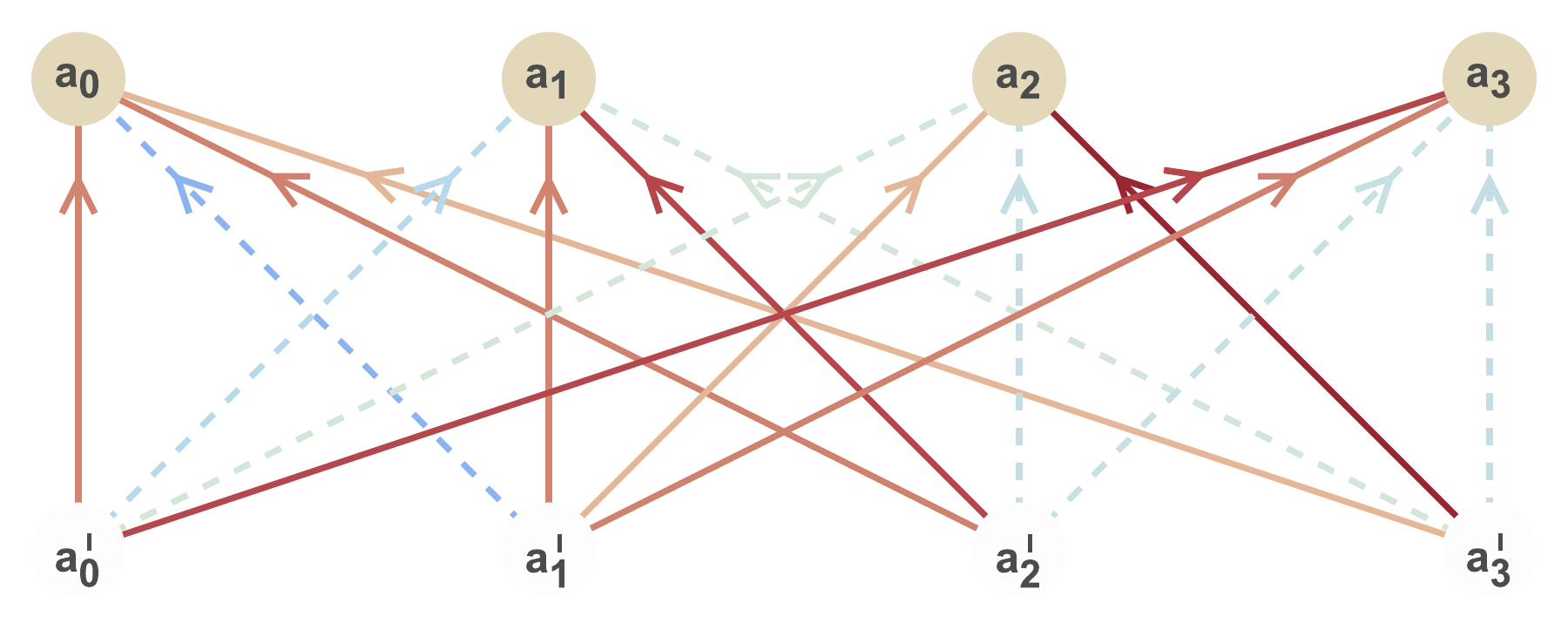}
         \caption{$S^{\hat{\mathcal{U}}_{\text{eg}}}_\textbf{DW}$ for $\mathcal{U}_{\text{eg}}[\bullet]=U_{\text{eg}} \bullet U_{\text{eg}}^\dagger $}
\end{subfigure} 
\begin{subfigure}{0.45\textwidth}
\centering
         \includegraphics[width=0.93\linewidth]{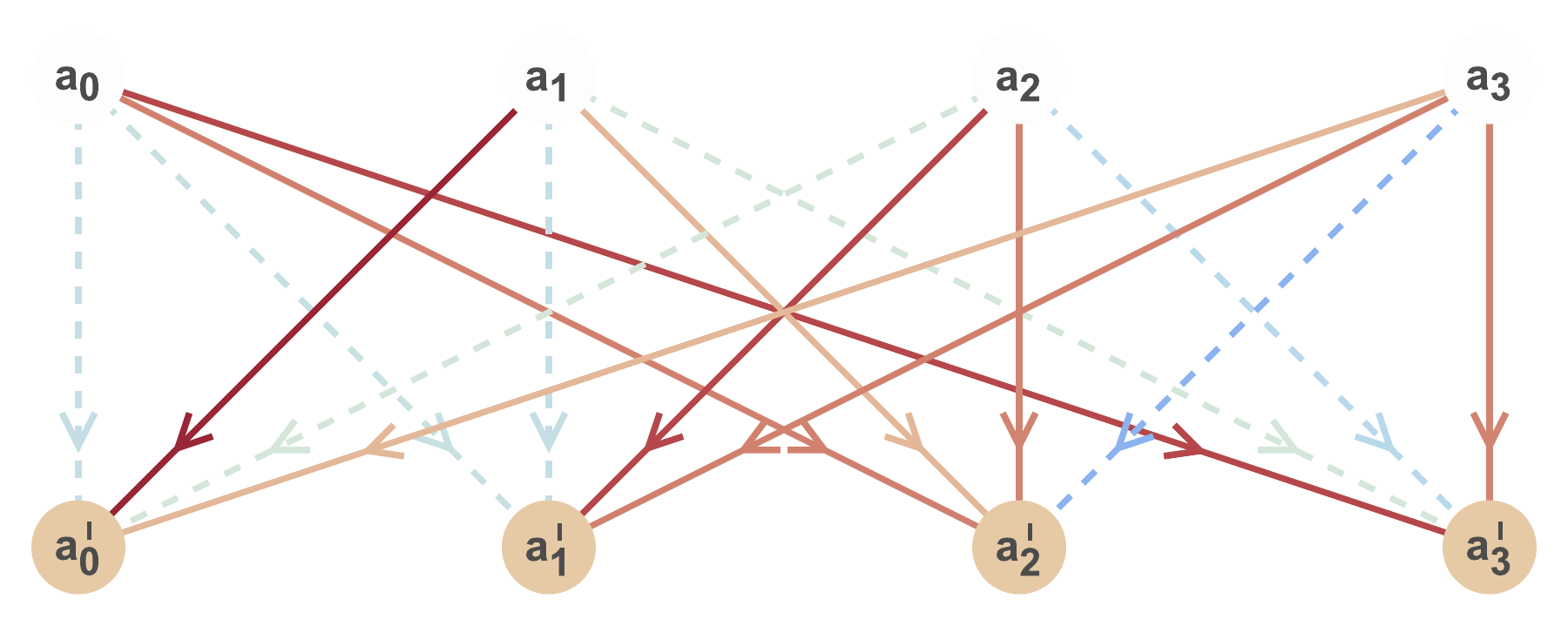}
         \caption{$S^{\mathcal{U}_{\text{eg}}}_\textbf{SP}$ for $\mathcal{U}_{\text{eg}}[\bullet]=U_{\text{eg}} \bullet U_{\text{eg}}^\dagger $}
\end{subfigure} 
\begin{subfigure}{0.45\textwidth}
\centering
         \includegraphics[width=0.93\linewidth]{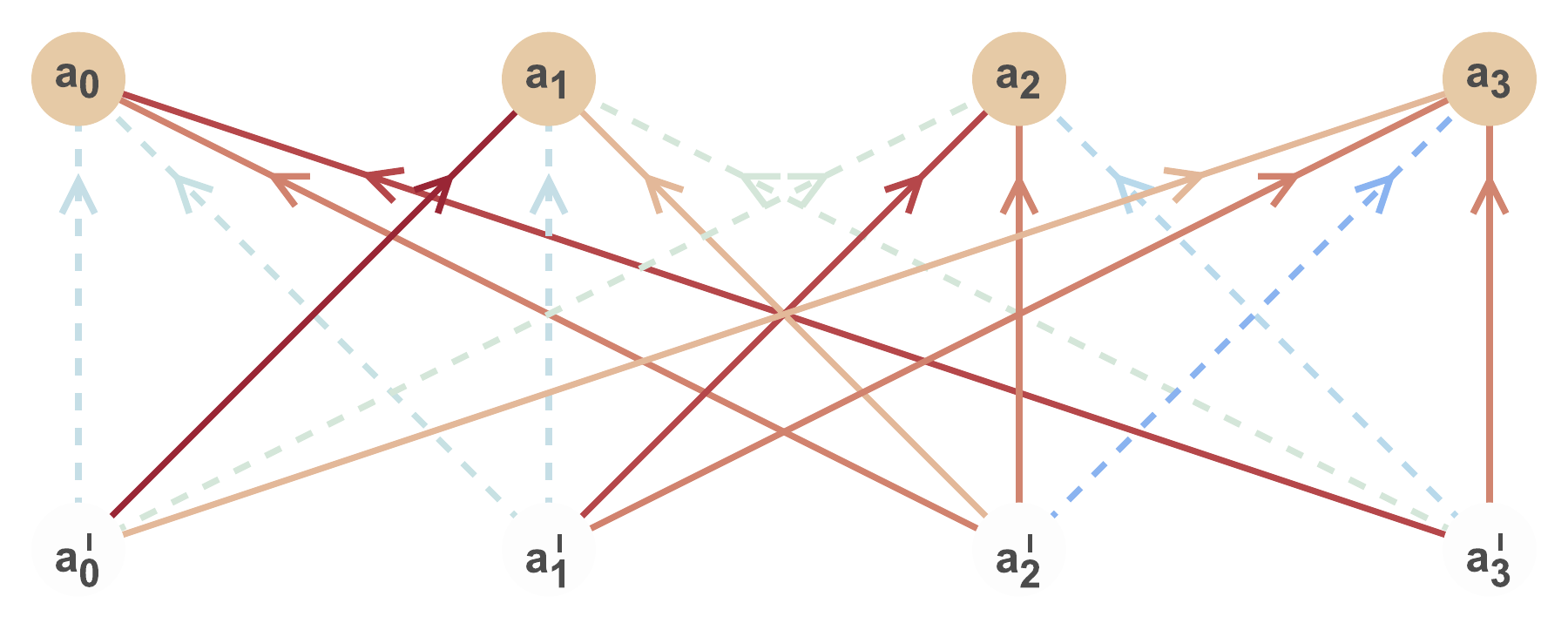}
         \caption{$S^{\hat{\mathcal{U}}_{\text{eg}}}_\textbf{SP}$ for $\mathcal{U}_{\text{eg}}[\bullet]=U_{\text{eg}} \bullet U_{\text{eg}}^\dagger $}
\end{subfigure} 
     \caption{Transition Graphs for the Hadamard gate and an arbitrarily chosen qubit unitary and their respective retrodictions.}
     \label{fig:hadaarbi}
\end{figure}

Likewise, an arbitrarily chosen unitary $U_{\text{eg}}$:
\begin{eqnarray*}
     U_{\text{eg}} \, &\hat{=}& \, \frac{1}{4} \begin{psmallmatrix}
 i \left(\sqrt{3}+2 i\right) & 0 & 3 i & 0 \\
 0 & i \left(\sqrt{3}+2 i\right) & 0 & 3 i \\
 -3 i & 0 & 2+i \sqrt{3} & 0 \\
 0 & -3 i & 0 & 2+i \sqrt{3} \\
     \end{psmallmatrix} \\
\end{eqnarray*}
has the following quasiprobability objects: 
\begin{equation*}
S^{\mathcal{U}_{\text{eg}}}_\textbf{DW} = \frac{1}{16} 
\begin{psmallmatrix}
 9 & \sqrt{3}-6 & 4-3 \sqrt{3} & 2 \sqrt{3}+9 \\
 -\sqrt{3}-6 & 9 & 9-2 \sqrt{3} & 3 \sqrt{3}+4 \\
 3 \sqrt{3}+4 & 2 \sqrt{3}+9 & -3 & 6-5 \sqrt{3} \\
 9-2 \sqrt{3} & 4-3 \sqrt{3} & 5 \sqrt{3}+6 & -3 \\ 
\end{psmallmatrix}
\end{equation*}
\begin{equation*}
S^{\mathcal{U}_{\text{eg}}}_\textbf{SP} = \frac{1}{16} 
\begin{psmallmatrix}
 -3 & 5 \sqrt{3}+6 & 4-3 \sqrt{3} & 9-2 \sqrt{3} \\
 6-5 \sqrt{3} & -3 & 2 \sqrt{3}+9 & 3 \sqrt{3}+4 \\
 3 \sqrt{3}+4 & 9-2 \sqrt{3} & 9 & -\sqrt{3}-6 \\
 2 \sqrt{3}+9 & 4-3 \sqrt{3} & \sqrt{3}-6 & 9 \\
\end{psmallmatrix}
\end{equation*}
It is clear that these forward channels do not give permutative QPRs. Nevertheless the property that $S^{\hat{\mathcal{U}}_\rf} = S^{\hat{\mathcal{U}}} = (S^{\mathcal{U}})^\text{T}$ is still reflected clearly in FIG. \ref{fig:hadaarbi}. In contrast to these reversible maps, we can speak of the quantum total erasure channel mentioned in section \ref{fullyrevfullirrev}. The full swap is expressed as such: 
\begin{eqnarray}\label{swapeq}
\swap \; \hat{=} \; 
\begin{psmallmatrix}
 1 & 0 & 0 & 0 \\
 0 & 0 & 1 & 0 \\
 0 & 1 & 0 & 0 \\
 0 & 0 & 0 & 1 \\
\end{psmallmatrix}\end{eqnarray}

As depicted clearly in FIG. \ref{fig:SWAP}, both the forward channel and its retrodiction are erase perfectly to the relevant state (the ancilla for the forward map and the reference state for the retrodiction). 

\clearpage 
\bibliography{refsret}

\begin{thebibliography}{43}%
\makeatletter
\providecommand \@ifxundefined [1]{%
 \@ifx{#1\undefined}
}%
\providecommand \@ifnum [1]{%
 \ifnum #1\expandafter \@firstoftwo
 \else \expandafter \@secondoftwo
 \fi
}%
\providecommand \@ifx [1]{%
 \ifx #1\expandafter \@firstoftwo
 \else \expandafter \@secondoftwo
 \fi
}%
\providecommand \natexlab [1]{#1}%
\providecommand \enquote  [1]{``#1''}%
\providecommand \bibnamefont  [1]{#1}%
\providecommand \bibfnamefont [1]{#1}%
\providecommand \citenamefont [1]{#1}%
\providecommand \href@noop [0]{\@secondoftwo}%
\providecommand \href [0]{\begingroup \@sanitize@url \@href}%
\providecommand \@href[1]{\@@startlink{#1}\@@href}%
\providecommand \@@href[1]{\endgroup#1\@@endlink}%
\providecommand \@sanitize@url [0]{\catcode `\\12\catcode `\$12\catcode
  `\&12\catcode `\#12\catcode `\^12\catcode `\_12\catcode `\%12\relax}%
\providecommand \@@startlink[1]{}%
\providecommand \@@endlink[0]{}%
\providecommand \url  [0]{\begingroup\@sanitize@url \@url }%
\providecommand \@url [1]{\endgroup\@href {#1}{\urlprefix }}%
\providecommand \urlprefix  [0]{URL }%
\providecommand \Eprint [0]{\href }%
\providecommand \doibase [0]{https://doi.org/}%
\providecommand \selectlanguage [0]{\@gobble}%
\providecommand \bibinfo  [0]{\@secondoftwo}%
\providecommand \bibfield  [0]{\@secondoftwo}%
\providecommand \translation [1]{[#1]}%
\providecommand \BibitemOpen [0]{}%
\providecommand \bibitemStop [0]{}%
\providecommand \bibitemNoStop [0]{.\EOS\space}%
\providecommand \EOS [0]{\spacefactor3000\relax}%
\providecommand \BibitemShut  [1]{\csname bibitem#1\endcsname}%
\let\auto@bib@innerbib\@empty
\bibitem [{\citenamefont {Watanabe}(1965)}]{watanabe65}%
  \BibitemOpen
  \bibfield  {author} {\bibinfo {author} {\bibfnamefont {S.}~\bibnamefont
  {Watanabe}},\ }\bibfield  {title} {\bibinfo {title} {Conditional
  probabilities in physics},\ }\href
  {https://doi.org/https://doi.org/10.1143/PTPS.E65.135} {\bibfield  {journal}
  {\bibinfo  {journal} {Progr. Theor. Phys. Suppl.}\ }\textbf {\bibinfo
  {volume} {E65}},\ \bibinfo {pages} {135} (\bibinfo {year}
  {1965})}\BibitemShut {NoStop}%
\bibitem [{\citenamefont {Watanabe}(1955)}]{watanabe55}%
  \BibitemOpen
  \bibfield  {author} {\bibinfo {author} {\bibfnamefont {S.}~\bibnamefont
  {Watanabe}},\ }\bibfield  {title} {\bibinfo {title} {Symmetry of physical
  laws. part iii. prediction and retrodiction},\ }\href
  {https://doi.org/10.1103/RevModPhys.27.179} {\bibfield  {journal} {\bibinfo
  {journal} {Rev. Mod. Phys.}\ }\textbf {\bibinfo {volume} {27}},\ \bibinfo
  {pages} {179} (\bibinfo {year} {1955})}\BibitemShut {NoStop}%
\bibitem [{\citenamefont {{Jeffrey}}(1965)}]{jeffrey}%
  \BibitemOpen
  \bibfield  {author} {\bibinfo {author} {\bibfnamefont {R.}~\bibnamefont
  {{Jeffrey}}},\ }\href@noop {} {\emph {\bibinfo {title} {The logic of
  decision}}}\ (\bibinfo  {publisher} {McGraw-Hill},\ \bibinfo {year}
  {1965})\BibitemShut {NoStop}%
\bibitem [{\citenamefont {Jaynes}(2003)}]{jaynes_2003}%
  \BibitemOpen
  \bibfield  {author} {\bibinfo {author} {\bibfnamefont {E.~T.}\ \bibnamefont
  {Jaynes}},\ }\href {https://doi.org/10.1017/CBO9780511790423} {\emph
  {\bibinfo {title} {Probability Theory: The Logic of Science}}}\ (\bibinfo
  {publisher} {Cambridge University Press},\ \bibinfo {year}
  {2003})\BibitemShut {NoStop}%
\bibitem [{\citenamefont {Parzygnat}\ and\ \citenamefont
  {Fullwood}(2022)}]{PF22}%
  \BibitemOpen
  \bibfield  {author} {\bibinfo {author} {\bibfnamefont {A.~J.}\ \bibnamefont
  {Parzygnat}}\ and\ \bibinfo {author} {\bibfnamefont {J.}~\bibnamefont
  {Fullwood}},\ }\href@noop {} {\bibinfo {title} {From time-reversal symmetry
  to quantum bayes' rules}} (\bibinfo {year} {2022}),\ \Eprint
  {https://arxiv.org/abs/2212.08088} {arXiv:2212.08088 [quant-ph]} \BibitemShut
  {NoStop}%
\bibitem [{\citenamefont {Petz}(1986)}]{petz1}%
  \BibitemOpen
  \bibfield  {author} {\bibinfo {author} {\bibfnamefont {D.}~\bibnamefont
  {Petz}},\ }\bibfield  {title} {\bibinfo {title} {Sufficient subalgebras and
  the relative entropy of states of a von neumann algebra},\ }\href
  {https://doi.org/10.1007/BF01212345} {\bibfield  {journal} {\bibinfo
  {journal} {Comm. Math. Phys.}\ }\textbf {\bibinfo {volume} {105}},\ \bibinfo
  {pages} {123} (\bibinfo {year} {1986})}\BibitemShut {NoStop}%
\bibitem [{\citenamefont {Petz}(1988)}]{petz}%
  \BibitemOpen
  \bibfield  {author} {\bibinfo {author} {\bibfnamefont {D.}~\bibnamefont
  {Petz}},\ }\bibfield  {title} {\bibinfo {title} {{Sufficiency of channels
  over von Neumann algebras}},\ }\href {https://doi.org/10.1093/qmath/39.1.97}
  {\bibfield  {journal} {\bibinfo  {journal} {The Quarterly Journal of
  Mathematics}\ }\textbf {\bibinfo {volume} {39}},\ \bibinfo {pages} {97}
  (\bibinfo {year} {1988})}\BibitemShut {NoStop}%
\bibitem [{\citenamefont {Wilde}(2015)}]{wilde-recov}%
  \BibitemOpen
  \bibfield  {author} {\bibinfo {author} {\bibfnamefont {M.}~\bibnamefont
  {Wilde}},\ }\bibfield  {title} {\bibinfo {title} {Recoverability in quantum
  information theory},\ }\href {https://doi.org/10.1098/rspa.2015.0338}
  {\bibfield  {journal} {\bibinfo  {journal} {Proceedings of the Royal Society
  A}\ }\textbf {\bibinfo {volume} {471}},\ \bibinfo {pages} {20150338}
  (\bibinfo {year} {2015})}\BibitemShut {NoStop}%
\bibitem [{\citenamefont {Wilde}(2011)}]{wilde2011CItoQIreview}%
  \BibitemOpen
  \bibfield  {author} {\bibinfo {author} {\bibfnamefont {M.~M.}\ \bibnamefont
  {Wilde}},\ }\bibfield  {title} {\bibinfo {title} {From classical to quantum
  shannon theory},\ }\href@noop {} {\bibfield  {journal} {\bibinfo  {journal}
  {arXiv preprint arXiv:1106.1445}\ } (\bibinfo {year} {2011})}\BibitemShut
  {NoStop}%
\bibitem [{\citenamefont {Li}\ and\ \citenamefont {Winter}(2018)}]{li-winter}%
  \BibitemOpen
  \bibfield  {author} {\bibinfo {author} {\bibfnamefont {K.}~\bibnamefont
  {Li}}\ and\ \bibinfo {author} {\bibfnamefont {A.}~\bibnamefont {Winter}},\
  }\bibfield  {title} {\bibinfo {title} {Squashed entanglement,
  k-extendibility, quantum markov chains, and recovery maps},\ }\href
  {https://doi.org/10.1007/s10701-018-0143-6} {\bibfield  {journal} {\bibinfo
  {journal} {Found. Phys.}\ }\textbf {\bibinfo {volume} {48}},\ \bibinfo
  {pages} {910} (\bibinfo {year} {2018})}\BibitemShut {NoStop}%
\bibitem [{\citenamefont {Leifer}\ and\ \citenamefont
  {Spekkens}(2013)}]{Leifer-Spekkens}%
  \BibitemOpen
  \bibfield  {author} {\bibinfo {author} {\bibfnamefont {M.~S.}\ \bibnamefont
  {Leifer}}\ and\ \bibinfo {author} {\bibfnamefont {R.~W.}\ \bibnamefont
  {Spekkens}},\ }\bibfield  {title} {\bibinfo {title} {Towards a formulation of
  quantum theory as a causally neutral theory of bayesian inference},\ }\href
  {https://doi.org/10.1103/PhysRevA.88.052130} {\bibfield  {journal} {\bibinfo
  {journal} {Phys. Rev. A}\ }\textbf {\bibinfo {volume} {88}},\ \bibinfo
  {pages} {052130} (\bibinfo {year} {2013})}\BibitemShut {NoStop}%
\bibitem [{\citenamefont {Parzygnat}\ and\ \citenamefont
  {Buscemi}(2022{\natexlab{a}})}]{petzisking2022axioms}%
  \BibitemOpen
  \bibfield  {author} {\bibinfo {author} {\bibfnamefont {A.~J.}\ \bibnamefont
  {Parzygnat}}\ and\ \bibinfo {author} {\bibfnamefont {F.}~\bibnamefont
  {Buscemi}},\ }\bibfield  {title} {\bibinfo {title} {Axioms for retrodiction:
  achieving time-reversal symmetry with a prior},\ }\href@noop {} {\bibfield
  {journal} {\bibinfo  {journal} {arXiv preprint arXiv:2210.13531}\ } (\bibinfo
  {year} {2022}{\natexlab{a}})}\BibitemShut {NoStop}%
\bibitem [{\citenamefont {Kwon}\ and\ \citenamefont {Kim}(2019)}]{kwon-kim}%
  \BibitemOpen
  \bibfield  {author} {\bibinfo {author} {\bibfnamefont {H.}~\bibnamefont
  {Kwon}}\ and\ \bibinfo {author} {\bibfnamefont {M.~S.}\ \bibnamefont {Kim}},\
  }\bibfield  {title} {\bibinfo {title} {Fluctuation theorems for a quantum
  channel},\ }\href {https://doi.org/10.1103/PhysRevX.9.031029} {\bibfield
  {journal} {\bibinfo  {journal} {Phys. Rev. X}\ }\textbf {\bibinfo {volume}
  {9}},\ \bibinfo {pages} {031029} (\bibinfo {year} {2019})}\BibitemShut
  {NoStop}%
\bibitem [{\citenamefont {Buscemi}\ and\ \citenamefont {Scarani}(2021)}]{BS21}%
  \BibitemOpen
  \bibfield  {author} {\bibinfo {author} {\bibfnamefont {F.}~\bibnamefont
  {Buscemi}}\ and\ \bibinfo {author} {\bibfnamefont {V.}~\bibnamefont
  {Scarani}},\ }\bibfield  {title} {\bibinfo {title} {Fluctuation theorems from
  bayesian retrodiction},\ }\href {https://doi.org/10.1103/PhysRevE.103.052111}
  {\bibfield  {journal} {\bibinfo  {journal} {Phys. Rev. E}\ }\textbf {\bibinfo
  {volume} {103}},\ \bibinfo {pages} {052111} (\bibinfo {year}
  {2021})}\BibitemShut {NoStop}%
\bibitem [{\citenamefont {Aw}\ \emph {et~al.}(2021)\citenamefont {Aw},
  \citenamefont {Buscemi},\ and\ \citenamefont {Scarani}}]{AwBS}%
  \BibitemOpen
  \bibfield  {author} {\bibinfo {author} {\bibfnamefont {C.~C.}\ \bibnamefont
  {Aw}}, \bibinfo {author} {\bibfnamefont {F.}~\bibnamefont {Buscemi}},\ and\
  \bibinfo {author} {\bibfnamefont {V.}~\bibnamefont {Scarani}},\ }\bibfield
  {title} {\bibinfo {title} {Fluctuation theorems with retrodiction rather than
  reverse processes},\ }\href {https://doi.org/10.1116/5.0060893} {\bibfield
  {journal} {\bibinfo  {journal} {AVS Quantum Science}\ }\textbf {\bibinfo
  {volume} {3}},\ \bibinfo {pages} {045601} (\bibinfo {year} {2021})},\ \Eprint
  {https://arxiv.org/abs/https://doi.org/10.1116/5.0060893}
  {https://doi.org/10.1116/5.0060893} \BibitemShut {NoStop}%
\bibitem [{\citenamefont {Ferrie}\ and\ \citenamefont
  {Emerson}(2009)}]{ferrie2009framed}%
  \BibitemOpen
  \bibfield  {author} {\bibinfo {author} {\bibfnamefont {C.}~\bibnamefont
  {Ferrie}}\ and\ \bibinfo {author} {\bibfnamefont {J.}~\bibnamefont
  {Emerson}},\ }\bibfield  {title} {\bibinfo {title} {Framed hilbert space:
  hanging the quasi-probability pictures of quantum theory},\ }\href
  {https://doi.org/10.1088/1367-2630/11/6/063040} {\bibfield  {journal}
  {\bibinfo  {journal} {New Journal of Physics}\ }\textbf {\bibinfo {volume}
  {11}},\ \bibinfo {pages} {063040} (\bibinfo {year} {2009})}\BibitemShut
  {NoStop}%
\bibitem [{\citenamefont {Ferrie}(2011)}]{ferrie2011quasi}%
  \BibitemOpen
  \bibfield  {author} {\bibinfo {author} {\bibfnamefont {C.}~\bibnamefont
  {Ferrie}},\ }\bibfield  {title} {\bibinfo {title} {Quasi-probability
  representations of quantum theory with applications to quantum information
  science},\ }\href {https://doi.org/10.1088/0034-4885/74/11/116001} {\bibfield
   {journal} {\bibinfo  {journal} {Reports on Progress in Physics}\ }\textbf
  {\bibinfo {volume} {74}},\ \bibinfo {pages} {116001} (\bibinfo {year}
  {2011})}\BibitemShut {NoStop}%
\bibitem [{\citenamefont {Ferrie}\ and\ \citenamefont
  {Emerson}(2008)}]{ferrie2008frame}%
  \BibitemOpen
  \bibfield  {author} {\bibinfo {author} {\bibfnamefont {C.}~\bibnamefont
  {Ferrie}}\ and\ \bibinfo {author} {\bibfnamefont {J.}~\bibnamefont
  {Emerson}},\ }\bibfield  {title} {\bibinfo {title} {Frame representations of
  quantum mechanics and the necessity of negativity in quasi-probability
  representations},\ }\href {https://doi.org/10.1088/1751-8113/41/35/352001}
  {\bibfield  {journal} {\bibinfo  {journal} {Journal of Physics A:
  Mathematical and Theoretical}\ }\textbf {\bibinfo {volume} {41}},\ \bibinfo
  {pages} {352001} (\bibinfo {year} {2008})}\BibitemShut {NoStop}%
\bibitem [{\citenamefont {Veitch}\ \emph {et~al.}(2012)\citenamefont {Veitch},
  \citenamefont {Ferrie}, \citenamefont {Gross},\ and\ \citenamefont
  {Emerson}}]{veitch2012negative}%
  \BibitemOpen
  \bibfield  {author} {\bibinfo {author} {\bibfnamefont {V.}~\bibnamefont
  {Veitch}}, \bibinfo {author} {\bibfnamefont {C.}~\bibnamefont {Ferrie}},
  \bibinfo {author} {\bibfnamefont {D.}~\bibnamefont {Gross}},\ and\ \bibinfo
  {author} {\bibfnamefont {J.}~\bibnamefont {Emerson}},\ }\bibfield  {title}
  {\bibinfo {title} {Negative quasi-probability as a resource for quantum
  computation},\ }\href@noop {} {\bibfield  {journal} {\bibinfo  {journal} {New
  Journal of Physics}\ }\textbf {\bibinfo {volume} {14}},\ \bibinfo {pages}
  {113011} (\bibinfo {year} {2012})}\BibitemShut {NoStop}%
\bibitem [{\citenamefont {Howard}\ \emph {et~al.}(2014)\citenamefont {Howard},
  \citenamefont {Wallman}, \citenamefont {Veitch},\ and\ \citenamefont
  {Emerson}}]{howard2014contextuality}%
  \BibitemOpen
  \bibfield  {author} {\bibinfo {author} {\bibfnamefont {M.}~\bibnamefont
  {Howard}}, \bibinfo {author} {\bibfnamefont {J.}~\bibnamefont {Wallman}},
  \bibinfo {author} {\bibfnamefont {V.}~\bibnamefont {Veitch}},\ and\ \bibinfo
  {author} {\bibfnamefont {J.}~\bibnamefont {Emerson}},\ }\bibfield  {title}
  {\bibinfo {title} {Contextuality supplies the ‘magic’for quantum
  computation},\ }\href {https://doi.org/https://doi.org/10.1038/nature13460}
  {\bibfield  {journal} {\bibinfo  {journal} {Nature}\ }\textbf {\bibinfo
  {volume} {510}},\ \bibinfo {pages} {351} (\bibinfo {year}
  {2014})}\BibitemShut {NoStop}%
\bibitem [{\citenamefont {Pashayan}\ \emph {et~al.}(2015)\citenamefont
  {Pashayan}, \citenamefont {Wallman},\ and\ \citenamefont
  {Bartlett}}]{pashayan2015estimating}%
  \BibitemOpen
  \bibfield  {author} {\bibinfo {author} {\bibfnamefont {H.}~\bibnamefont
  {Pashayan}}, \bibinfo {author} {\bibfnamefont {J.~J.}\ \bibnamefont
  {Wallman}},\ and\ \bibinfo {author} {\bibfnamefont {S.~D.}\ \bibnamefont
  {Bartlett}},\ }\bibfield  {title} {\bibinfo {title} {Estimating outcome
  probabilities of quantum circuits using quasiprobabilities},\ }\href
  {https://doi.org/10.1103/PhysRevLett.115.070501} {\bibfield  {journal}
  {\bibinfo  {journal} {Phys. Rev. Lett.}\ }\textbf {\bibinfo {volume} {115}},\
  \bibinfo {pages} {070501} (\bibinfo {year} {2015})}\BibitemShut {NoStop}%
\bibitem [{Note1()}]{Note1}%
  \BibitemOpen
  \bibinfo {note} {Of course, one can have it that $a'$ and $a$ are defined in
  different state spaces $A$ and $A'$, but we can always take $A'' = A \cup A'$
  and characterize the channel in this larger alphabet.}\BibitemShut {Stop}%
\bibitem [{Note2()}]{Note2}%
  \BibitemOpen
  \bibinfo {note} {This constraint also exists in the classical Bayes update
  and is likewise of no practical concern as one can always ensure that that
  $\gamma $ is full-rank by adding some arbitrarily small weights into its
  spectrum and adding some arbitrarily small mapping probability in $\protect
  \mathcal {E}$ as well. These contributions can then be sent to zero on the
  recovered state.}\BibitemShut {Stop}%
\bibitem [{\citenamefont {Parzygnat}\ and\ \citenamefont
  {Buscemi}(2022{\natexlab{b}})}]{PB22}%
  \BibitemOpen
  \bibfield  {author} {\bibinfo {author} {\bibfnamefont {A.~J.}\ \bibnamefont
  {Parzygnat}}\ and\ \bibinfo {author} {\bibfnamefont {F.}~\bibnamefont
  {Buscemi}},\ }\href@noop {} {\bibinfo {title} {Axioms for retrodiction:
  achieving time-reversal symmetry with a prior}} (\bibinfo {year}
  {2022}{\natexlab{b}}),\ \Eprint {https://arxiv.org/abs/2210.13531}
  {arXiv:2210.13531 [quant-ph]} \BibitemShut {NoStop}%
\bibitem [{\citenamefont {Mac~Lane}(1998)}]{mac1998categories}%
  \BibitemOpen
  \bibfield  {author} {\bibinfo {author} {\bibfnamefont {S.}~\bibnamefont
  {Mac~Lane}},\ }\href@noop {} {\emph {\bibinfo {title} {Categories for the
  Working Mathematician}}},\ Vol.~\bibinfo {volume} {5}\ (\bibinfo  {publisher}
  {Springer Science \& Business Media},\ \bibinfo {year} {1998})\ pp.\ \bibinfo
  {pages} {13--30}\BibitemShut {NoStop}%
\bibitem [{\citenamefont {Carnap}(2002)}]{carnap2002logical}%
  \BibitemOpen
  \bibfield  {author} {\bibinfo {author} {\bibfnamefont {R.}~\bibnamefont
  {Carnap}},\ }\href@noop {} {\emph {\bibinfo {title} {The logical syntax of
  language}}}\ (\bibinfo  {publisher} {Open Court Publishing},\ \bibinfo {year}
  {2002})\BibitemShut {NoStop}%
\bibitem [{Note3()}]{Note3}%
  \BibitemOpen
  \bibinfo {note} {It is the case, for SIC-POVM and Discrete Wigner
  representation, that $\protect \text {\protect \normalfont {Tr}}\protect
  \!\left [F_j\right ] = 1/d$ for all $j \in \Lambda $. But this is not
  generally the case for all valid QPRs.}\BibitemShut {Stop}%
\bibitem [{\citenamefont {Ruzzi}\ and\ \citenamefont
  {Galetti}(2000)}]{ruzzi2000quantum}%
  \BibitemOpen
  \bibfield  {author} {\bibinfo {author} {\bibfnamefont {M.}~\bibnamefont
  {Ruzzi}}\ and\ \bibinfo {author} {\bibfnamefont {D.}~\bibnamefont
  {Galetti}},\ }\bibfield  {title} {\bibinfo {title} {Quantum discrete phase
  space dynamics and its continuous limit},\ }\href
  {https://doi.org/10.1088/0305-4470/33/5/317} {\bibfield  {journal} {\bibinfo
  {journal} {Journal of Physics A: Mathematical and General}\ }\textbf
  {\bibinfo {volume} {33}},\ \bibinfo {pages} {1065} (\bibinfo {year}
  {2000})}\BibitemShut {NoStop}%
\bibitem [{\citenamefont {Zhu}(2016)}]{zhu2016quasiprobability}%
  \BibitemOpen
  \bibfield  {author} {\bibinfo {author} {\bibfnamefont {H.}~\bibnamefont
  {Zhu}},\ }\bibfield  {title} {\bibinfo {title} {Quasiprobability
  representations of quantum mechanics with minimal negativity},\ }\href
  {https://doi.org/10.1103/PhysRevLett.117.120404} {\bibfield  {journal}
  {\bibinfo  {journal} {Phys. Rev. Lett.}\ }\textbf {\bibinfo {volume} {117}},\
  \bibinfo {pages} {120404} (\bibinfo {year} {2016})}\BibitemShut {NoStop}%
\bibitem [{\citenamefont {Wootters}(1987)}]{wootters1987wigner}%
  \BibitemOpen
  \bibfield  {author} {\bibinfo {author} {\bibfnamefont {W.~K.}\ \bibnamefont
  {Wootters}},\ }\bibfield  {title} {\bibinfo {title} {A wigner-function
  formulation of finite-state quantum mechanics},\ }\href
  {https://doi.org/https://doi.org/10.1016/0003-4916(87)90176-X} {\bibfield
  {journal} {\bibinfo  {journal} {Annals of Physics}\ }\textbf {\bibinfo
  {volume} {176}},\ \bibinfo {pages} {1} (\bibinfo {year} {1987})}\BibitemShut
  {NoStop}%
\bibitem [{\citenamefont {Klimov}\ and\ \citenamefont
  {Muñoz}(2005)}]{klimov2005discrete}%
  \BibitemOpen
  \bibfield  {author} {\bibinfo {author} {\bibfnamefont {A.~B.}\ \bibnamefont
  {Klimov}}\ and\ \bibinfo {author} {\bibfnamefont {C.}~\bibnamefont
  {Muñoz}},\ }\bibfield  {title} {\bibinfo {title} {Discrete wigner function
  dynamics},\ }\href {https://doi.org/10.1088/1464-4266/7/12/022} {\bibfield
  {journal} {\bibinfo  {journal} {Journal of Optics B: Quantum and
  Semiclassical Optics}\ }\textbf {\bibinfo {volume} {7}},\ \bibinfo {pages}
  {S588} (\bibinfo {year} {2005})}\BibitemShut {NoStop}%
\bibitem [{\citenamefont {Gibbons}\ \emph {et~al.}(2004)\citenamefont
  {Gibbons}, \citenamefont {Hoffman},\ and\ \citenamefont
  {Wootters}}]{gibbons2004discrete}%
  \BibitemOpen
  \bibfield  {author} {\bibinfo {author} {\bibfnamefont {K.~S.}\ \bibnamefont
  {Gibbons}}, \bibinfo {author} {\bibfnamefont {M.~J.}\ \bibnamefont
  {Hoffman}},\ and\ \bibinfo {author} {\bibfnamefont {W.~K.}\ \bibnamefont
  {Wootters}},\ }\bibfield  {title} {\bibinfo {title} {Discrete phase space
  based on finite fields},\ }\href {https://doi.org/10.1103/PhysRevA.70.062101}
  {\bibfield  {journal} {\bibinfo  {journal} {Phys. Rev. A}\ }\textbf {\bibinfo
  {volume} {70}},\ \bibinfo {pages} {062101} (\bibinfo {year}
  {2004})}\BibitemShut {NoStop}%
\bibitem [{\citenamefont {Gross}(2006)}]{gross2006hudson}%
  \BibitemOpen
  \bibfield  {author} {\bibinfo {author} {\bibfnamefont {D.}~\bibnamefont
  {Gross}},\ }\bibfield  {title} {\bibinfo {title} {Hudson’s theorem for
  finite-dimensional quantum systems},\ }\href
  {https://doi.org/10.1063/1.2393152} {\bibfield  {journal} {\bibinfo
  {journal} {Journal of Mathematical Physics}\ }\textbf {\bibinfo {volume}
  {47}},\ \bibinfo {pages} {122107} (\bibinfo {year} {2006})},\ \Eprint
  {https://arxiv.org/abs/https://doi.org/10.1063/1.2393152}
  {https://doi.org/10.1063/1.2393152} \BibitemShut {NoStop}%
\bibitem [{\citenamefont {Appleby}\ \emph {et~al.}(2017)\citenamefont
  {Appleby}, \citenamefont {Fuchs}, \citenamefont {Stacey},\ and\ \citenamefont
  {Zhu}}]{appleby2017introducing}%
  \BibitemOpen
  \bibfield  {author} {\bibinfo {author} {\bibfnamefont {M.}~\bibnamefont
  {Appleby}}, \bibinfo {author} {\bibfnamefont {C.~A.}\ \bibnamefont {Fuchs}},
  \bibinfo {author} {\bibfnamefont {B.~C.}\ \bibnamefont {Stacey}},\ and\
  \bibinfo {author} {\bibfnamefont {H.}~\bibnamefont {Zhu}},\ }\bibfield
  {title} {\bibinfo {title} {Introducing the qplex: a novel arena for quantum
  theory},\ }\href {https://doi.org/10.1140/epjd/e2017-80024-y} {\bibfield
  {journal} {\bibinfo  {journal} {The European Physical Journal D}\ }\textbf
  {\bibinfo {volume} {71}},\ \bibinfo {pages} {1} (\bibinfo {year}
  {2017})}\BibitemShut {NoStop}%
\bibitem [{\citenamefont {Kiktenko}\ \emph {et~al.}(2020)\citenamefont
  {Kiktenko}, \citenamefont {Malyshev}, \citenamefont {Mastiukova},
  \citenamefont {Man'ko}, \citenamefont {Fedorov},\ and\ \citenamefont
  {Chru\ifmmode \acute{s}\else \'{s}\fi{}ci\ifmmode~\acute{n}\else
  \'{n}\fi{}ski}}]{kiktenko2020probability}%
  \BibitemOpen
  \bibfield  {author} {\bibinfo {author} {\bibfnamefont {E.~O.}\ \bibnamefont
  {Kiktenko}}, \bibinfo {author} {\bibfnamefont {A.~O.}\ \bibnamefont
  {Malyshev}}, \bibinfo {author} {\bibfnamefont {A.~S.}\ \bibnamefont
  {Mastiukova}}, \bibinfo {author} {\bibfnamefont {V.~I.}\ \bibnamefont
  {Man'ko}}, \bibinfo {author} {\bibfnamefont {A.~K.}\ \bibnamefont
  {Fedorov}},\ and\ \bibinfo {author} {\bibfnamefont {D.}~\bibnamefont
  {Chru\ifmmode \acute{s}\else \'{s}\fi{}ci\ifmmode~\acute{n}\else
  \'{n}\fi{}ski}},\ }\bibfield  {title} {\bibinfo {title} {Probability
  representation of quantum dynamics using pseudostochastic maps},\ }\href
  {https://doi.org/10.1103/PhysRevA.101.052320} {\bibfield  {journal} {\bibinfo
   {journal} {Phys. Rev. A}\ }\textbf {\bibinfo {volume} {101}},\ \bibinfo
  {pages} {052320} (\bibinfo {year} {2020})}\BibitemShut {NoStop}%
\bibitem [{\citenamefont {DeBrota}\ \emph {et~al.}()\citenamefont {DeBrota},
  \citenamefont {Fuchs},\ and\ \citenamefont {Stacey}}]{debrota_qbism}%
  \BibitemOpen
  \bibfield  {author} {\bibinfo {author} {\bibfnamefont {J.}~\bibnamefont
  {DeBrota}}, \bibinfo {author} {\bibfnamefont {C.}~\bibnamefont {Fuchs}},\
  and\ \bibinfo {author} {\bibfnamefont {B.}~\bibnamefont {Stacey}},\ }\href
  {http://www.physics.umb.edu/Research/QBism/} {\bibinfo {title} {Qbism
  research group}},\ \bibinfo {howpublished}
  {\url{http://www.physics.umb.edu/Research/QBism/}}\BibitemShut {NoStop}%
\bibitem [{\citenamefont {Appleby}\ \emph {et~al.}(2013)\citenamefont
  {Appleby}, \citenamefont {Yadsan-Appleby},\ and\ \citenamefont
  {Zauner}}]{appleby2013galois}%
  \BibitemOpen
  \bibfield  {author} {\bibinfo {author} {\bibfnamefont {D.~M.}\ \bibnamefont
  {Appleby}}, \bibinfo {author} {\bibfnamefont {H.}~\bibnamefont
  {Yadsan-Appleby}},\ and\ \bibinfo {author} {\bibfnamefont {G.}~\bibnamefont
  {Zauner}},\ }\bibfield  {title} {\bibinfo {title} {Galois automorphisms of a
  symmetric measurement},\ }\href@noop {} {\bibfield  {journal} {\bibinfo
  {journal} {Quantum Info. Comput.}\ }\textbf {\bibinfo {volume} {13}},\
  \bibinfo {pages} {672–720} (\bibinfo {year} {2013})}\BibitemShut {NoStop}%
\bibitem [{\citenamefont {Braasch}\ and\ \citenamefont
  {Wootters}(2020)}]{braasch2020transition}%
  \BibitemOpen
  \bibfield  {author} {\bibinfo {author} {\bibfnamefont {W.~F.}\ \bibnamefont
  {Braasch}}\ and\ \bibinfo {author} {\bibfnamefont {W.~K.}\ \bibnamefont
  {Wootters}},\ }\bibfield  {title} {\bibinfo {title} {Transition probabilities
  and transition rates in discrete phase space},\ }\href
  {https://doi.org/10.1103/PhysRevA.102.052204} {\bibfield  {journal} {\bibinfo
   {journal} {Phys. Rev. A}\ }\textbf {\bibinfo {volume} {102}},\ \bibinfo
  {pages} {052204} (\bibinfo {year} {2020})}\BibitemShut {NoStop}%
\bibitem [{Note4()}]{Note4}%
  \BibitemOpen
  \bibinfo {note} {{Our main claim, summarized in Table \ref {tab:compares}, is
  that, in deriving the retrodiction \protect \textit {inside a QPR}, the
  structure coefficients are used only to obtain the $X_\gamma $ from the
  $v_\gamma $. Eq.~\protect \eqref {eq:altse} shows that one can use the
  structure coefficients to obtain the QPR $S^\protect \mathcal {E}$ \protect
  \textit {from the objects of the Hilbert space}: thus, this fact does not
  contradict our main claim.}}\BibitemShut {Stop}%
\bibitem [{Note5()}]{Note5}%
  \BibitemOpen
  \bibinfo {note} {For NQPR and SIC-POVMs, the proof is even simpler using the
  fact that $\DOTSB \sum@ \slimits@ _x G_x = d \protect \mathbb {1}$: indeed,
  in this case we have $\DOTSB \sum@ \slimits@ _{xy} \protect \text {\protect
  \normalfont {Tr}}\protect \!\left [F_i G_x G_j G_y\right ]=d^2 \protect \text
  {\protect \normalfont {Tr}}\protect \!\left [F_i G_j\right ]$; but $\DOTSB
  \sum@ \slimits@ _{xy}\delta _{ix} \delta _{jy} \delta _{ij}=\delta _{ij}$,
  and therefore \protect \eqref {claxprej} can hold only if $\hskip 1em\relax
  d^2 \protect \text {\protect \normalfont {Tr}}\protect \!\left [F_i G_j\right
  ] = \delta _{ij}$ i.e.~$d^2=1$.}\BibitemShut {Stop}%
\bibitem [{Note6()}]{Note6}%
  \BibitemOpen
  \bibinfo {note} {The equivalence in colour is a statement of irreversibility
  since it implies $\protect \mathcal {E}(a'|a) = \protect \hat {\gamma }(a')$.
  Thus $\DOTSB \sum@ \slimits@ _{a}\protect \mathcal {E}(a'|a)q(a) = \protect
  \hat {\gamma }(a')$ for all $q(a)$, which is just to say the channel
  irreversibily erases all information about the input.}\BibitemShut {Stop}%
\bibitem [{\citenamefont {Arvidsson-Shukur}\ \emph {et~al.}(2020)\citenamefont
  {Arvidsson-Shukur}, \citenamefont {Yunger~Halpern}, \citenamefont {Lepage},
  \citenamefont {Lasek}, \citenamefont {Barnes},\ and\ \citenamefont
  {Lloyd}}]{arv2020}%
  \BibitemOpen
  \bibfield  {author} {\bibinfo {author} {\bibfnamefont {D.~R.~M.}\
  \bibnamefont {Arvidsson-Shukur}}, \bibinfo {author} {\bibfnamefont
  {N.}~\bibnamefont {Yunger~Halpern}}, \bibinfo {author} {\bibfnamefont
  {H.~V.}\ \bibnamefont {Lepage}}, \bibinfo {author} {\bibfnamefont {A.~A.}\
  \bibnamefont {Lasek}}, \bibinfo {author} {\bibfnamefont {C.~H.~W.}\
  \bibnamefont {Barnes}},\ and\ \bibinfo {author} {\bibfnamefont
  {S.}~\bibnamefont {Lloyd}},\ }\bibfield  {title} {\bibinfo {title} {Quantum
  advantage in postselected metrology},\ }\href
  {https://doi.org/10.1038/s41467-020-17559-w} {\bibfield  {journal} {\bibinfo
  {journal} {Nature Communications}\ }\textbf {\bibinfo {volume} {11}},\
  \bibinfo {pages} {3775} (\bibinfo {year} {2020})}\BibitemShut {NoStop}%
\bibitem [{\citenamefont {Renes}\ \emph {et~al.}(2004)\citenamefont {Renes},
  \citenamefont {Blume-Kohout}, \citenamefont {Scott},\ and\ \citenamefont
  {Caves}}]{renes2004symmetric}%
  \BibitemOpen
  \bibfield  {author} {\bibinfo {author} {\bibfnamefont {J.~M.}\ \bibnamefont
  {Renes}}, \bibinfo {author} {\bibfnamefont {R.}~\bibnamefont {Blume-Kohout}},
  \bibinfo {author} {\bibfnamefont {A.~J.}\ \bibnamefont {Scott}},\ and\
  \bibinfo {author} {\bibfnamefont {C.~M.}\ \bibnamefont {Caves}},\ }\bibfield
  {title} {\bibinfo {title} {Symmetric informationally complete quantum
  measurements},\ }\href {https://doi.org/10.1063/1.1737053} {\bibfield
  {journal} {\bibinfo  {journal} {Journal of Mathematical Physics}\ }\textbf
  {\bibinfo {volume} {45}},\ \bibinfo {pages} {2171} (\bibinfo {year}
  {2004})}\BibitemShut {NoStop}%
\end{thebibliography}%

\end{document}